\documentclass[journal=jacsat,manuscript=article]{achemso}

\usepackage[utf8]{inputenc}
\usepackage{graphicx}
\usepackage{amsmath}
\usepackage{newtxmath}
\usepackage{color}
\usepackage{multirow}
\usepackage{hyperref}
\usepackage{breqn}
\usepackage{physics}

\usepackage{nameref}

\usepackage{threeparttable}
\usepackage{subfiles}
\usepackage{tabularx}
\usepackage{array}
\usepackage{booktabs}
\usepackage{float}
\usepackage{longtable}
\usepackage{threeparttablex}
\usepackage{makecell}
\usepackage{adjustbox}

\author{Giorgio Visentin}
\affiliation{Department of Physics and Astronomy,
University College London,
London WC1E 6BT
United Kingdom}
\email{ucapgvi@ucl.ac.uk}

\author{Agapi Emmanouilidou}
\affiliation{Department of Physics and Astronomy,
University College London,
London WC1E 6BT
United Kingdom}
\email{a.emmanouilidou@ucl.ac.uk}

\title{Potential-energy surfaces of water and its molecular ions up to \texorpdfstring{H$_2$O$^{3+}$}{H2O3+}}

\begin{document}

\begin{tocentry}
\hspace*{0.3cm}%
\raisebox{-2.95cm}{%
\includegraphics[
width=13.47cm,
height=7.5cm
]{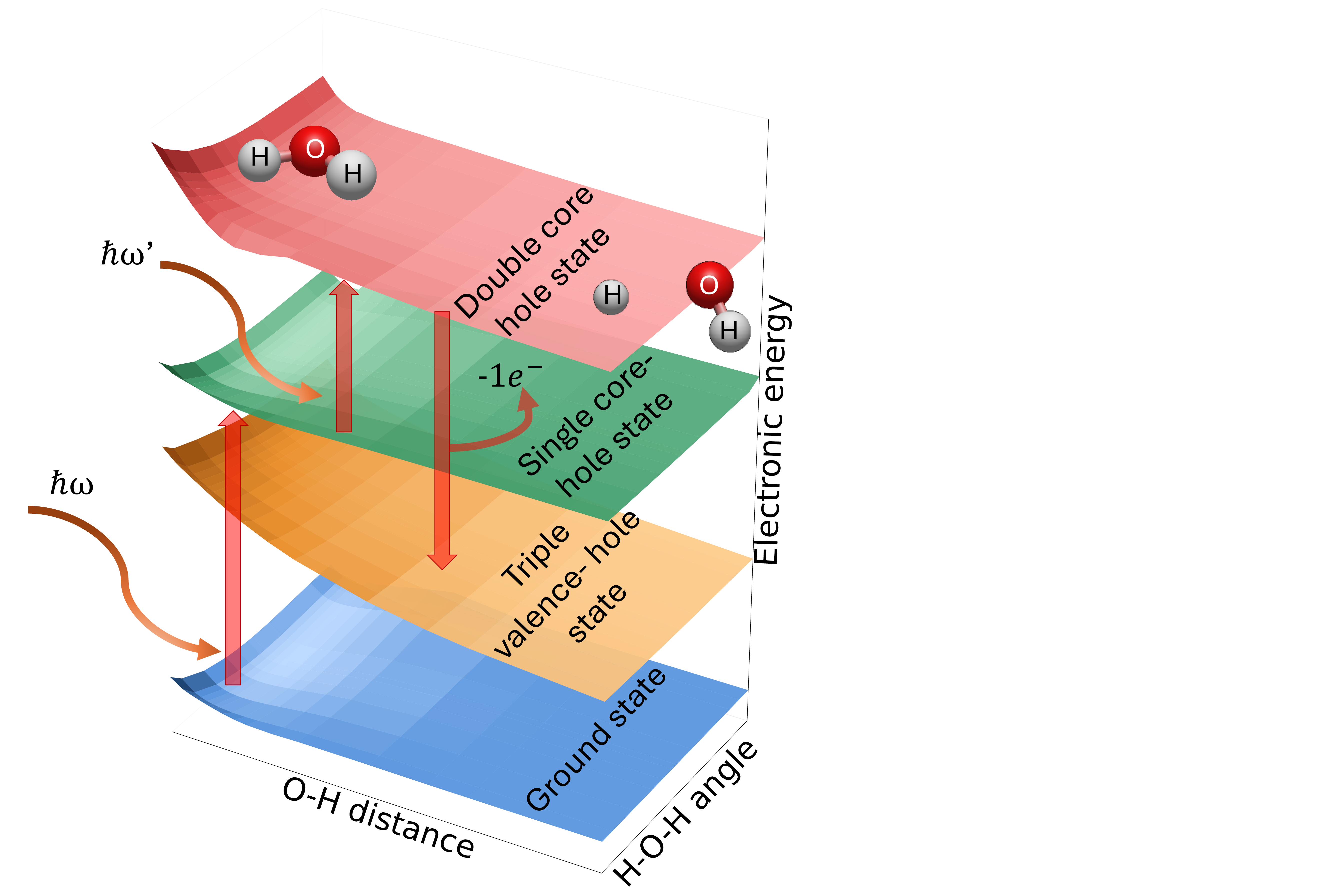}%
}
\end{tocentry}

\begin{abstract}
    We compute the potential-energy surfaces for molecular ions up to \texorpdfstring{H$_2$O$^{3+}$}{H2O3+} with any combinations of outer-valence, inner-valence and core holes. To obtain the potential-energy surfaces, we consider several~\textit{ab initio} 
 techniques
 based on the CASSCF method and benchmark them with respect to the literature. We thus identify the
  techniques
 which produce accurate potential-energy surfaces as a function of the two bond lengths and the angle between the two bond lengths. These potential-energy surfaces  will be used in future studies  of the interaction of water with free-electron laser pulses. Furthermore, the techniques employed in this work in the context of water can be used to obtain the potential-energy surfaces of other triatomic molecules.
\end{abstract}
\date{\today}
\maketitle
\section{Introduction}
Extensive research has been devoted to molecules driven by  intense laser pulses. The interaction of molecules with x-rays is central to a wide range of applications, ranging from medical imaging and therapy~\cite{Breitkreutz:2020} to the manipulation of materials~\cite{Cavalieri:2007,Li:2007}. 
%
When x-rays interact with a molecule, electrons from core molecular orbitals can escape to the continuum by single-photon absorption. This escape and the subsequent Auger-Meitner decay~\cite{Meitner:1922, Li:2022, Haynes:2021,Ismail:2024} that occurs  in molecular ions with core vacancies involve the interplay of electron and nuclear motion~\cite{Travnikova:2016, Carniato:2025}. Hence, the simulation of these ultrafast molecular processes should account for the coupling of  electron and nuclear motion \cite{Bhattacharya:2021,Hadjipittas:2023, Mountney:2025,Calegari:2014}.

One way to include the nuclear motion is to account for the force between the nuclei by  computing  potential-energy surfaces (PESs)~\cite{Bhattacharya:2021,Hadjipittas:2023, Mountney:2025,Calegari:2014}
and then to substitute this force in classical equations of motion~\cite{Mountney:2024,Wang:2026}.  However, reasonably accurate computations of PESs associated with single or multiple core-hole states are not straightforward, due to the difficulty of  \textit{ab initio} calculations converging  for states lying high in energy~\cite{Hadjipittas:2023,Bhattacharya:2021}. Computing the PESs for triatomic molecules is further complicated by the three degrees of freedom necessary to account for the motion of the nuclei, instead of just the  internuclear distance relevant to diatomic molecules. In this work, we compute the PESs of water  
up to triply charged ions. We take the three degrees of freedom to be  the two bond lengths of O-H and the angle between them.
%
Accurate computations of the PESs are particularly timely as recent experimental studies ~\cite{Ismail:2024,Wang:2023,Jahnke:2021,Sankari:2020,Ertan:2018} have demonstrated the importance of nuclear motion in the interaction of water with X-rays. 

Concerning water, most theoretical studies involve the computation of PESs for ionic states with electrons missing from valence orbitals~\cite{Leclerc:1974,Smith:1975, Dehareng:1983,Nobusada:2000}. However, theoretical studies of water ions with core electrons missing do not involve the computation of PESs but rather  the accurate calculation  of ionization potentials \cite{Gervais:2009, Otsuka:2002, Ida:2008, Tarantelli:1985, DeMoura:2022,Streit:2009,Mucke:2013,Handke:1995},
an exception being the (1a$_1)^{-1}$ $^2$A$_1$ state~\cite{Sankari:2003}.
The main reason for the scarcity of studies on the computation of PESs for core-hole water ions and other molecular ions in general is  variational collapse~\cite{Besley:2009}.
Among the methods employed in the theoretical  studies focusing  on ionization potentials are the 
single-reference configuration-interaction (CI)~\cite{Leclerc:1974}
and multi-reference configuration-interaction (MRCI)~\cite{Gervais:2009,Cesar:1989,Handke:1995,Sankari:2003},  density functional theory~\cite{Otsuka:2002}, Diffusion Quantum Monte Carlo \cite{Streit:2009}, two-electron Dyson propagator  \cite{Ida:2008}, and Green's function based methods \cite{Tarantelli:1985, DeMoura:2022}. 
For triply charged core- and valence-hole water ions, ionization potentials are computed by  employing the complete-active-space self-consistent-field (CASSCF) method \cite{Mucke:2013} and  the MRCI and Green's function-based second-order algebraic diagrammatic construction approximation method (ADC(2))  \cite{Handke:1995}.  

%
Here, addressing the scarcity  of PESs for water ions, we systematically compute
the potential-energy surfaces of the singly, doubly and triply charged water ions with any combination of core- and valence-holes. We compute these PESs employing the \textit{ab initio} CASSCF method in the context of the quantum-chemistry package MOLPRO~\cite{Molpro_1, Molpro_2}. Namely, our work extends to water ions the \textit{ab initio} approaches developed for the  N$_2$ ion states by Bhattacharya~\textit{et al.}~\cite{Bhattacharya:2021} and Hadjipittas~\textit{et al.}~\cite{Hadjipittas:2023}. We compute accurate PESs for up to triply charged water ions employing techniques that can be generalized to other triatomic molecules. 
 Future studies may employ these PESs to account for nuclear motion in the interaction of water with X-ray pulses.
The \textit{ab initio}  approaches we employ to compute the PESs of water ions are benchmarked for ionization potentials with respect to measurements~\cite{Siegbahn:1974,Siegbahn:1975,Mucke:2013}, and computations~\cite{Otsuka:2002,Mucke:2013,Cesar:1989,DeMoura:2022,Gervais:2009,Streit:2009,Ida:2008,Tarantelli:1985,Handke:1995} for the singly, doubly and  triply charged ions, for the latter for the ones   lowest in energy.%
\section{Theoretical background}
%
The symmetry group of water is C$_{2v}$~\cite{BunkerBOOK, BishopBOOK}. The ground state of water is referred to as $^1$A$_1$~\cite{Smith:1975}, with its electron configuration given by (1a$_1)^{2}$(2a$_1)^{2}$(1b$_2)^{2}$(3a$_1)^{2}$(1b$_1)^{2}$ for the water molecule being on the y-z plane \cite{Handke:1995}.  The core molecular orbitals are 1a$_1$ and 2a$_1$, where 1a$_1$ stems from  the 1s atomic orbital of oxygen, while 2a$_1$ stems from the linear combination of the 2s atomic orbital of oxygen  and  1s atomic orbitals of the two hydrogens \cite{Pople:1957}.  The  valence molecular orbitals are  1b$_2$,
and  3a$_1$ which stem from the linear combination of  a 2p orbital of O and the 1s orbitals of the two H atoms, as well as the 
non-bonding 1b$_1$ orbital which stems from a 2p orbital of  O \cite{Pople:1950}.
%
Here, we compute the PESs of water ions in terms of the two O-H distances, R$\mathrm{_{O-H}}$, and the H-O-H angle, $\thetaup$. The  radial grid varies from 0.7 \AA{} to 4.0 \AA{},  
with more points taken around the equilibrium distance of neutral  water equal to 0.956
\AA{}, while the angular grid varies from 30$^{\circ}$ to 180$^{\circ}$. 
To obtain the potential energies at points in between the grid points, we employ a  cubic-spline interpolation. Also, while we compute the PESs, for simplicity, we mostly plot cuts through these PESs, i.e.  we plot the resulting potential-energy curves (PECs) as a function of the bond angle or one bond distance. At the end of each section, we also plot PESs for two of the ion states considered in this section. 
%

In this work, we compute all   PESs for up to triply charged water ions. Specifically, we consider 
 \textit{single valence-hole} (SVH), \textit{double valence-hole} (DVH) and~\textit{triple valence-hole} (TVH) states and similarly   core-hole states, SCH, DCH and TCH, as well as    states with a combination of core and valence electrons missing with a charge of two, \textit{double core-valence-hole} (DCVH), and charge of three, 
  \textit{triple core-valence-hole} (TCVH) states.  
%
 We also note that the asymmetric stretching of the two bonds lowers the symmetry of water to the C$_s$ point group~\cite{Nobusada:2000}, where the only symmetry element is the plane of the molecule~\cite{BunkerBOOK, BishopBOOK}. Within this symmetry group, the 1a$_1$, 2a$_1$, 1b$_2$  and 3a$_1$ molecular orbitals turn into 1a$'$, 2a$'$, 3a$'$ and 4a$'$, respectively, while 1b$_1$ turns into 1a$''$. That is, in C$_s$,  the valence 1b$_2$ orbital becomes an inner-valence one, 3a$'$, since it has the same symmetry as the 4a$'$. The electronic configuration of water in C$_s$ is (1a$')^{2}$(2a$')^{2}$(3a$')^{2}$(4a$')^{2}$(1a$')^{2}$.
 Hence, to account for the asymmetric stretching of the two bonds, we compute the  orbitals within the C$_s$ group. For consistency with 
 the notation adopted in Molecular Spectroscopy~\cite{Aasbrink:1971,Page:1988}, in what follows,  we label the  water ions  according to C$_{2v}$.
  
To compute the molecular orbitals, we employ two different basis sets, namely,  aug-cc-pVQZ \cite{Dunning:1989,Kendall:1992}  for ion states with valence holes 
and  aug-cc-pCVQZ  \cite{Dunning:1989,Kendall:1992,Woon:1995} for states with at least one core hole.  This choice of the basis sets has been previously adopted by Bhattacharya~\textit{et al.}~\cite{Bhattacharya:2021} and Hadjipittas~\textit{et al.}~\cite{Hadjipittas:2023} for  N$_2$. 
To obtain  the molecular orbitals of the H$_2$O ions, we first obtain the  Hartree-Fock orbitals of the ground state of water. Then, we  perform  a CASSCF calculation on the ground state of water. The CASSCF 
computations are performed with two different active spaces depending on the water ion.  One active space includes eight molecular orbitals, all the ground-state molecular orbitals plus the virtual ones 4a$_1$ (5a$'$), 2b$_2$ (6a$'$) and 2b$_1$ (2a$''$), while the other active space includes an additional 5a$_1$ (7a$'$) virtual orbital; note that in parenthesis we denote the label of each orbital in the C$_s$ symmetry group. The latter active space is used to obtain the molecular orbitals for  the SCH, DCVH, DCH, TVH, TCH, TCVH water ions.  Following the work of Bhattacharya~\textit{et al.}~\cite{Bhattacharya:2021} and Hadjipittas~\textit{et al.}~\cite{Hadjipittas:2023}, these preliminary  molecular orbitals are the input to one of four different techniques that we employ to obtain the final molecular orbitals.   The \textit{ab initio} techniques and the basis sets employed depending on the water ions  are summarized in Table~\ref{Table:Approach} and discussed in detail in the following sections.
 The techniques we employ include   a single CASSCF optimization of the molecular orbitals,  state-averaging CASSCF (SA-CASSCF)~\cite{Granovski:2015},  and optimization of the core and valence CASSCF molecular orbitals in two different steps to avoid the variational collapse to the lowest energy state \cite{Bhattacharya:2021, Hadjipittas:2023, Besley:2009}. We refer to this latter  technique as TS-CASSCF~\cite{Rocha:2011,Rocha:2011-2,DeMoura:2013,Corral:2017,Carravetta:2013,Bhattacharya:2021,Hadjipittas:2023}. Another technique we employ  combines state averaging and two-step optimization within the CASSCF method, and we refer to it as SA-TS-CASSCF~\cite{Bhattacharya:2021,Hadjipittas:2023,Besley:2009}.
Mainly for the computation of ionization potentials, we also employ MRCI, starting from the CASSCF orbitals, for all the SVH and some of the DVH states to benchmark the accuracy of our computations against known results~\cite{Siegbahn:1974,Siegbahn:1975,Mucke:2013,Otsuka:2002,Mucke:2013,Cesar:1989,DeMoura:2022,Gervais:2009,Streit:2009,Ida:2008,Tarantelli:1985,Handke:1995}. 
Details on the computation of the PESs  depending on the water ion  are provided in the following sections.
\begin{table}[H]
\centering
\begin{adjustbox}{max width=\textwidth, max totalheight=1.2\textheight}
\begin{tabular}{c c | c c c c | c c | c c}
\hline
\hline
\multirow{2}{*}{Charge state} &
\multirow{2}{*}{Missing electrons} &
\multicolumn{4}{c|}{Technique} &
\multicolumn{2}{c|}{Basis set} &
\multicolumn{2}{c}{\# Molecular Orbitals in Active Space} \\
\\
\hline

& &
\multicolumn{1}{c}{CASSCF} &
\multicolumn{1}{c}{SA-CASSCF} &
\multicolumn{1}{c}{TS-CASSCF} &
\multicolumn{1}{c|}{SA-TS-CASSCF} &
\multicolumn{1}{c}{Aug-cc-pVQZ} &
\multicolumn{1}{c|}{Aug-cc-pCVQZ} &
8 &
9 \\
\hline
\\

\multirow{3}{*}{+1}
& Outer valence
& \checkmark & \checkmark & & & \checkmark & & \checkmark & \\
& Inner valence
& & \checkmark & & & \checkmark & & \checkmark & \\
& Core
& & & \checkmark & & & \checkmark & & \checkmark \\ \\

\multirow{6}{*}{+2}
& 2 Outer valence
& & \checkmark & & & \checkmark & & \checkmark & \\
& Outer + Inner valence
& & \checkmark & & & \checkmark & & \checkmark & \\
& 2 Inner valence
& & \checkmark & & & \checkmark & & \checkmark & \\
& Outer valence + Core
& &  & & \checkmark &  &\checkmark & &\checkmark \\
& Inner valence + Core
&  & & & \checkmark & & \checkmark & & \checkmark \\
& 2 Core
& & &\checkmark & \checkmark & & \checkmark & & \checkmark \\ \\

\multirow{6}{*}{+3}
& 3 Outer valence
& & \checkmark & & & \checkmark & & & \checkmark \\
& 2 Outer + Inner valence
& & \checkmark & & & \checkmark & & & \checkmark \\
& Outer + 2 Inner valence
& & \checkmark & & & \checkmark & & & \checkmark \\
& 2 Outer/Inner valence + Core
& & & & \checkmark & & \checkmark & & \checkmark \\
& Outer/Inner valence + 2 Core
& & & & \checkmark & & \checkmark & & \checkmark \\
& 3 Core
& & & & \checkmark & & \checkmark & & \checkmark \\

\hline
\hline
\end{tabular}
\end{adjustbox}
\caption{\textit{Ab initio} settings employed to compute the PESs of up to triply charged water ions.}
\label{Table:Approach}
\end{table}
\section{Results and discussion}
%
Using CASSCF and MRCI,  we find  the equilibrium bond distance of water to be 0.956~\AA{}, while the bond angle equals 104.51$^\circ$. These results are in excellent agreement with the results of Hoy~\textit{et al.}~\cite{Hoy:1979} and the more recent computations of Ho~\textit{et al.}~\cite{Ho:1996} and Partridge~\textit{et al.}~\cite{Partridge:1997}
In what follows, we present the PESs as a function of one bond length and the bond angle  as well as  the vertical ionization potentials (VIPs) of the ions with respect to neutral water. The VIPs are obtained by subtracting the energies of each ion from the energy of  neutral water both computed at  the  equilibrium geometry of water. 
\subsection{Singly charged ion states}
The PESs of the SVH ions are computed with the aug-cc-pVQZ basis set and the smaller active space, as shown in Table~\ref{Table:Approach}.
 We note that the techniques we employ to compute the PESs of the SVH ion states of water are the CASSCF for the outer-valence-hole state $\mathrm{(1b_{1})^{-1}}$, and the SA-CASSCF for the inner-valence
-hole state $\mathrm{(1b_{2})^{-1}}$, as for our previous studies for diatomic molecules in Refs \cite{Bhattacharya:2021,Hadjipittas:2023}. However, for the outer-valence-hold state $\mathrm{(3a_{1})^{-1}}$, instead
 of the CASSCF technique we would have employed for a diatomic molecule, we use the SA-CASSCF for water, since there is an avoided crossing between the same symmetry  (3a$_1)^{-1}$ $^2$A$_1$ and
 (1b$_2)^{-1}$ $^2$B$_2$ molecular states.
 
\begin{figure}[H]
    \centering
    \includegraphics[width=0.9\linewidth]{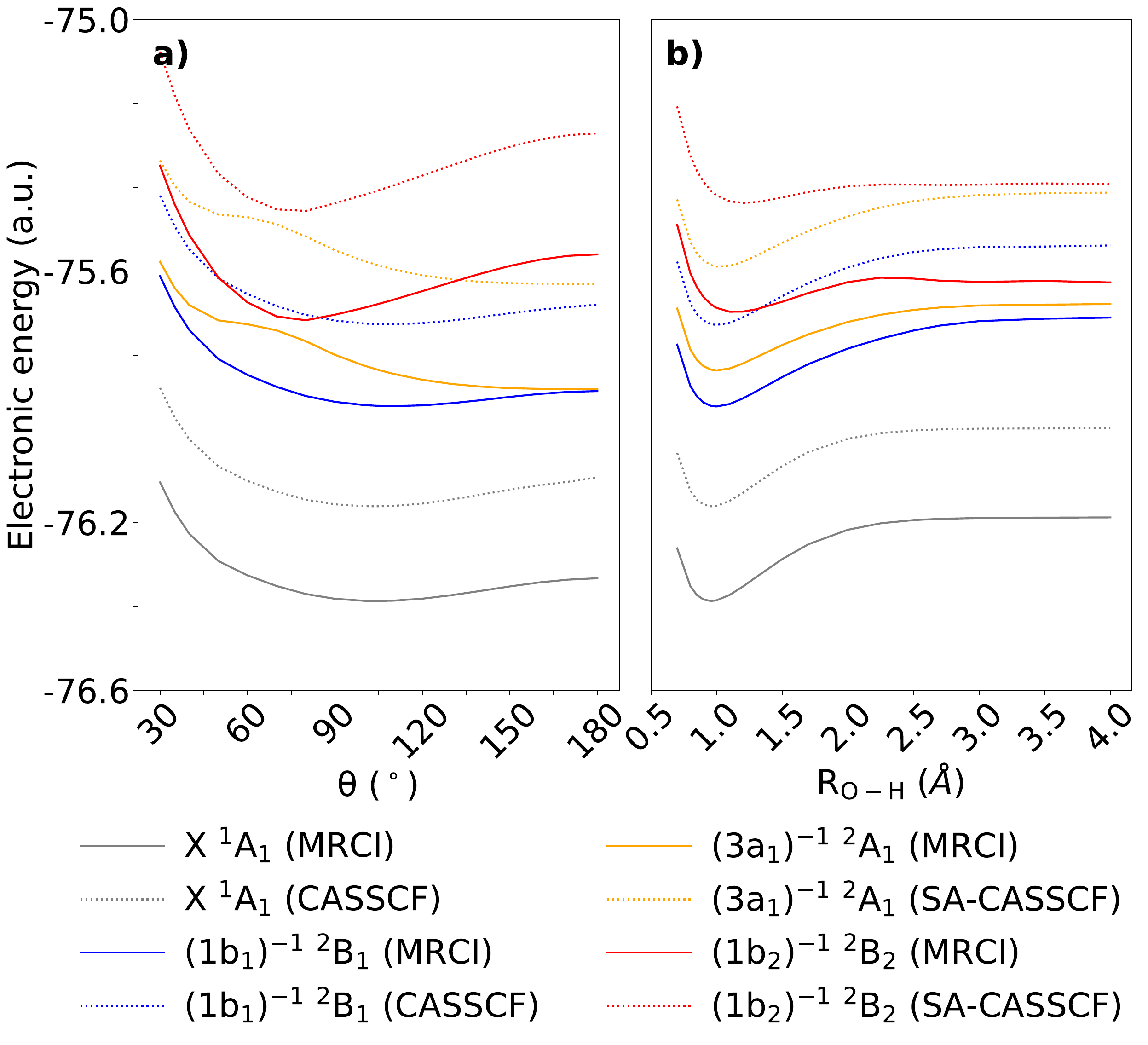}
    \caption{Potential-energy curves of the ground state and single-valence-hole  states of H$_2$O as a functions of a) the bond angle with both the bond distances frozen at the ground-state equilibrium geometry of neutral water and as a function of b) one of the bond distances, with the other bond distance and bond angle fixed the  ground-state equilibrium geometry of neutral water.  }
    \label{Fig:SVH_PECs}
\end{figure}

We first focus on the VIPs of the SVH states and benchmark them against the measurements from Electron Spectroscopy for Chemical Analysis (ESCA) of Siegbahn~\cite{Siegbahn:1974} and Density-Functional Theory (DFT) calculations by Otsuka~\textit{et al.}~\cite{Otsuka:2002}, as shown in Table~\ref{Table:SH}.
Our results are in very good agreement with the literature. As expected, MRCI provides more accurate results compared to CASSCF, that is, the VIPs computed with MRCI are closer to the values provided in the literature by both experiment and theory. 
\begin{table}[H]
\centering
\begin{threeparttable}

\begin{tabular}{c| c| c c}
\hline
\hline
\multirow{2}{*}{Electron configuration} & 
\multirow{2}{*}{State} & 
\multicolumn{2}{c}{Vertical Ionization Potential (eV)} \\
& & Our work & Literature \\
\hline

(1b$_1)^{-1}$ & $^2$B$_1$ & 
\makecell{12.62(MRCI) \\ 11.82(CASSCF)}& 
\makecell{12.62[a] \\ 12.72[b]} \\ \\

(3a$_1)^{-1}$ & $^2$A$_1$ & 
\makecell{14.97(MRCI) \\ 15.66(SA-CASSCF) \\ 13.89(CASSCF)} & 
\makecell{14.73[a] \\ 14.74[b]} \\ \\

(1b$_2)^{-1}$ & $^2$B$_2$ & 
\makecell{19.21(MRCI) \\ 20.48(SA-CASSCF)} & 
\makecell{18.55[a] \\ 19.25[b]} \\ \\

(2a$_1)^{-1}$  & $^2$A$_1$ & 
32.10(TS-CASSCF) & 
\makecell{32.12[a] \\ 31.84[b]} \\ \\

(1a$_1)^{-1}$  & $^2$A$_1$ & 
538.60(TS-CASSCF) & 
\makecell{539.93[a] \\ 539.90[b] \\ 539.86[c] \\ 539.65[d] \\ 538.53[e]}\\

\hline
\end{tabular}

\begin{tablenotes}[para]
\item \hfill\
\parbox{1.3\linewidth}{

\item[\textsuperscript{[a]}] ESCA measurements  \cite{Siegbahn:1974}
\item[\textsuperscript{[b]}]  DFT calculations  \cite{Otsuka:2002}
\item[\textsuperscript{[c]}] SA-CASSCF calculations  \cite{Mucke:2013}
\item[\textsuperscript{[d]}] MRCI calculations  \cite{Cesar:1989}
\item[\textsuperscript{[e]}] MR-ADC(2)-X calculations  \cite{DeMoura:2022}behavior
}
\hfill
\end{tablenotes}

\end{threeparttable}
\caption{Vertical ionization energies associated with the singly-ionized charge states of H$_2$O. We denote the states in terms of the electrons missing and their term symbols. The ions with one electron missing from the 1b$_1$ (1a$''$ in C$_s$), 3a$_1$ (4a$'$ in C$_s$) or 1b$_2$ (3a$'$ in C$_s$) orbitals are SVH ionic states, whereas the ions with one electron missing from either 2a$_1$ (2a$'$ in C$_s$) or 1a$_1$ (1a$'$ in C$_s$) orbitals are SCH ionic states.}
\label{Table:SH}
\end{table}
The PESs associated with the ground state of water and its SVH ions are shown in Figure~\ref{Fig:SVH_PECs}. The figure consists  of  two subplots with a) plotting the PESs along the bond angle ($\thetaup$), while the two O-H bond distances (R$\mathrm{_{O-H}}$) are fixed at the equilibrium geometry of neutral water. In Figure~\ref{Fig:SVH_PECs} b) the PESs are plotted along one R$\mathrm{_{O-H}}$, while the other R$\mathrm{_{O-H}}$ and the bond angle $\thetaup$ are fixed at the equilibrium geometry of neutral water. Throughout this work, as in Figure~\ref{Fig:SVH_PECs}, the PESs will be plotted as a function of one bond distance in one subplot and as a function of the bond angle in the other.
%
Figure~\ref{Fig:SVH_PECs} shows the potential-energy curves (PECs) obtained with CASSCF (dotted lines) and MRCI calculations (solid lines). We find that both sets of curves have the same behavior as a function of either the bond distance or the bond angle, while the MRCI curves are roughly 5 eV lower in energy.

\begin{figure}[H]
    \centering
    \includegraphics[width=0.9\linewidth]{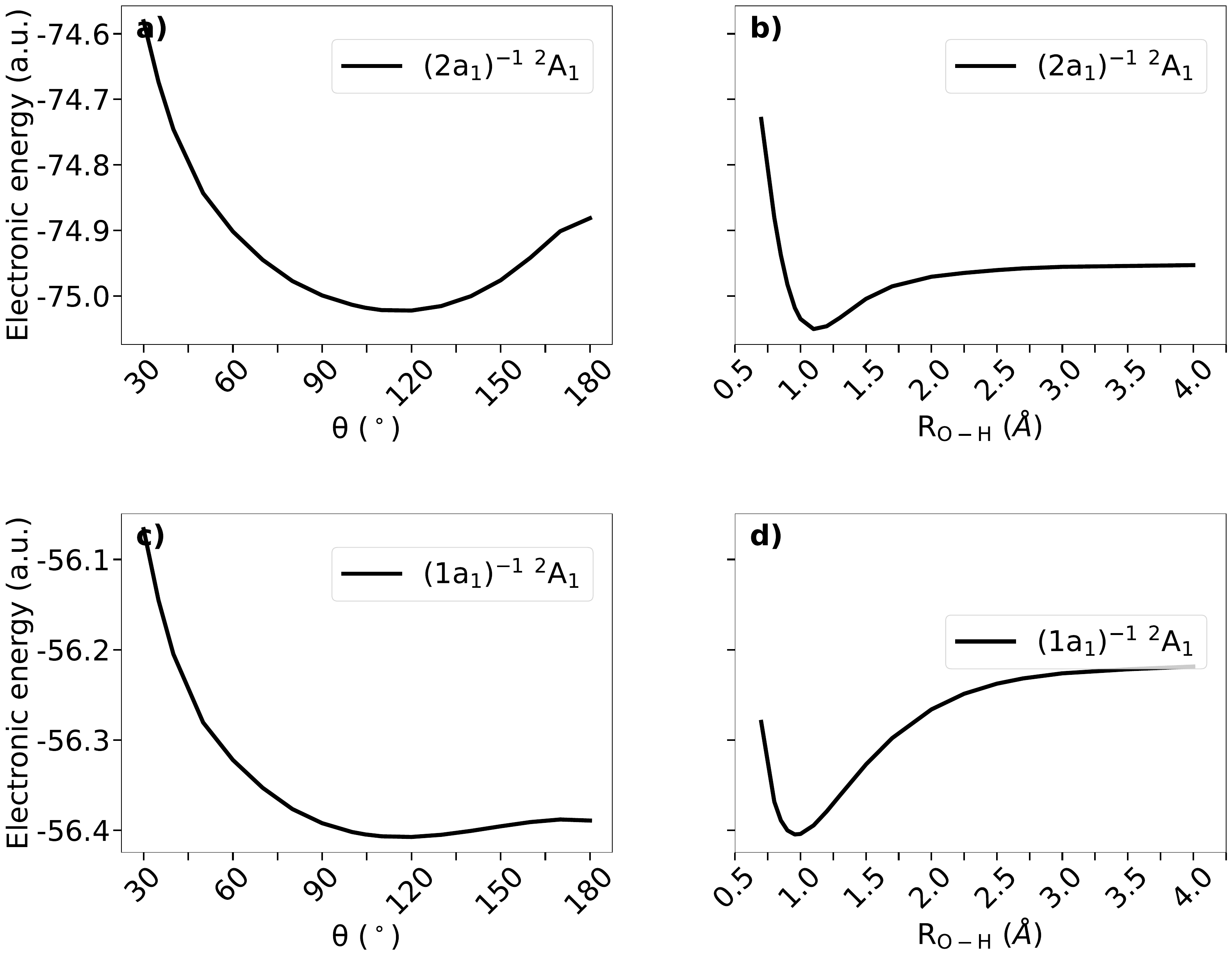}
    \caption{Potential-energy curves as in Figure~\ref{Fig:SVH_PECs} for the (2a$_1)^{-1}$ $^2$A$_1$ state (panels a) and b)) and the (1a$_1)^{-1}$ $^2$A$_1$ (panels c) and d)) state. }
    \label{Fig:SCH_PECs}
\end{figure}

\begin{figure}[H]
    \centering
    \includegraphics[width=0.9\linewidth]{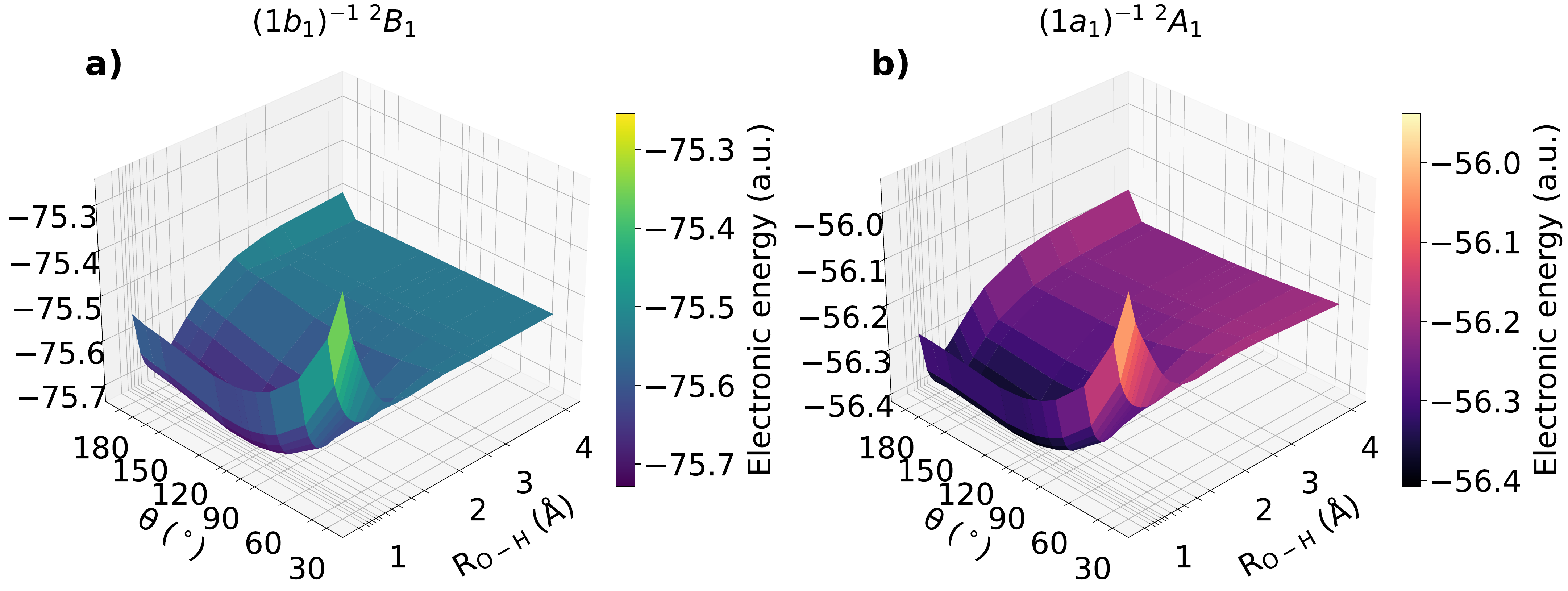}
    \caption{Potential-energy surfaces for the a) (1b$_1)^{-1}$ $^2$B$_1$ and  b) (1a$_1)^{-1}$ $^2$A$_1$ ion states of H$_2$O as functions of one distance and angle, while the other bond is fixed at the equilibrium bond distance of neutral water.}
    \label{Fig:SVH-SCH_PES}
\end{figure}

Another interesting feature in the PESs of water is the presence of avoided crossings that can occur for molecular states of the same symmetry. For water, the PESs are defined in terms of three degrees of freedom, hence avoided crossings can be present when plotting PECs as a function of one degree of freedom \cite{Lizuain:2009,Neumann,Naqvi:1972}. In Figure~\ref{Fig:SVH_PECs} a), we find an avoided crossing 
 between the (3a$_1)^{-1}$ $^2$A$_1$ and (1b$_2)^{-1}$ $^2$B$_2$ molecular states at around 70$^\circ$, with both these states having the same symmetry in the C$_s$ symmetry group. We also find that 
  the PECs of the (1b$_1)^{-1}$ $^2$B$_1$ and (3a$_1)^{-1}$ $^2$A$_1$ states have the same values for large bending angles approaching 180$^\circ$. Indeed, at this angle $\thetaup$ the molecule becomes linear and the molecular orbitals are described by the C$_{\infty v}$ group. The two orbitals 3a$_1$ and 1b$_1$ become $\pi_{u}$ orbitals of oxygen and are hence degenerate in energy.
 We also find that the PECs of SVH states have shallow minima as a function of the bond angle, while they have deeper minima as a function of the bond length. Concerning PECs as a function of the bond length, we find that the state (1b$_2)^{-1}$ $^2$B$_2$ features the shallowest minimum, whose depth amounts to 1.21 eV (CASSCF) and 1.90 eV (MRCI). This shallow minimum  is ascribed to the significant weakening of the O-H bond upon ionization of the bonding 1b$_2$ molecular orbital. 
 These trends also affect the PESs calculated by Leclerc~\textit{et al.}~\cite{Leclerc:1974} with the CI method. 
 %
 We show the PES for the SVH (1b$_1)^{-1}$ $^2$B$_1$ state computed with the CASSCF technique in Figure~\ref	{Fig:SVH-SCH_PES} a). The surface is plotted as a function of one bond distance and the bond angle, while the other bond distance is fixed to the equilibrium bond distance of neutral water. The PES exhibits the trends of the related cuts (PECs) in  Figure~\ref{Fig:SVH_PECs} a) and b), namely the shallow well along the bond angle  and the deeper one as a function of distance.
%
Next, we consider the  SCH states, which consist of the two $^2$A$_1$ states associated with the (1a$_1)^{-1}$ and (2a$_1)^{-1}$ electronic configurations. Both states are computed on the larger active space and the aug-cc-pCVQZ basis set, with the TS-CASSCF technique as summarized in Table~\ref{Table:Approach}.
We list VIPs of these core-hole states in Table~\ref{Table:SH} and compare them with measurements and computations from the literature. The computations in the literature are performed with CASSCF and MRCI by  Mucke~\textit{et al.}~\cite{Mucke:2013} and Cesar~\textit{et al.}~\cite{Cesar:1989}, respectively, the multi-reference algebraic-diagrammatic-construction-theory (MR-ADC(2)-X) of De Moura and Sokolov~\cite{DeMoura:2022} and the DFT technique by Otsuka~\textit{et al.}~\cite{Otsuka:2002}.
Our computations  are in very good agreement with the results from literature, suggesting  that the TS-CASSCF technique models these states with reasonable accuracy. In particular, the  VIP of 32.1 eV for the (2a$_1)^{-1}$ $^2$A$_1$ state deviates from the experiment by no more than 0.02 eV, while overestimating the calculations of Ref.~\cite{Otsuka:2002} by just 0.24 eV. The VIP of 538.6 eV for the (1a$_1)^{-1}$ $^2$A$_1$ state falls between the MR-ADC(2)-X calculations~\cite{DeMoura:2022} and the experimental data~\cite{Siegbahn:1974}, from which it differs by roughly 1.3 eV.
%
The PECs for the two core-hole  states are plotted in Figure~\ref{Fig:SCH_PECs}. In panels a) and b), the PEC for (2a$_1)^{-1}$ $^2$A$_1$ is plotted as a function of  $\thetaup$ and R$\mathrm{_{O-H}}$, respectively, while analogous plots are shown for the PEC of (1a$_1)^{-1}$ $^2$A$_1$ in panels c) and d). For both core-hole states, the potential-energy curves as a function of the bond length compared to  as a function of the bond angle have more pronounced well and minima. 
In particular, the CI (1a$_1)^{-1}$ $^2$A$_1$ PEC obtained along the bond angle of Sankari~\textit{et al.}~\cite{Sankari:2003} also shows a shallow minimum in the same range of bond angles as in our PES. 
 Furthermore, for this state, the same trends seen in the two panels of Figure~\ref{Fig:SCH_PECs} are visible in Figure~\ref{Fig:SVH-SCH_PES} b), where the PES is plotted in functions of one bond distance and the bond angle. 
\subsection{Doubly charged ion states}
We now focus on the doubly charged ions of water and their PESs. First, we discuss the lowest in energy doubly charged states, which consist of nine DVH ionic states of singlet and triplet spin symmetry. 
We compute the related PESs with the smaller active space and the aug-cc-pVQZ basis set, using the SA-CASSCF technique, see Table~\ref{Table:Approach}. Our choice of the SA-CASSCF technique is dictated by these states being within a narrow energy interval, as evident  by their VIPs in Table~\ref{Table:DCH}. 
We benchmark the accuracy of our  results for the VIPs by also performing MRCI calculations  for the closed-shell (1b$_1)^{-2}$ $^1$A$_1$ and (3a$_1)^{-2}$ $^1$A$_1$ states. Specifically, our computations are assessed against the MRCI calculations of Gervais~\textit{et al.}~\cite{Gervais:2009}, the Auger-spectroscopy measurements of Siegbahn~\textit{et al.}~\cite{Siegbahn:1975} and the electrostatic-electron spectroscopy measurements of Moddeman~\textit{et al.}~\cite{Moddeman:1971}. 
We find that our results  with SA-CASSCF and MRCI for the VIPS of these two DVH states are very similar.   Also, we find excellent agreement with the experimental results, while  our  results obtained with SA-CASSCF are close to the ones provided by  the MRCI calculations of Gervais~\textit{et al.}~\cite{Gervais:2009}. %

\begin{figure}
    \centering
    \includegraphics[width=0.9\linewidth]{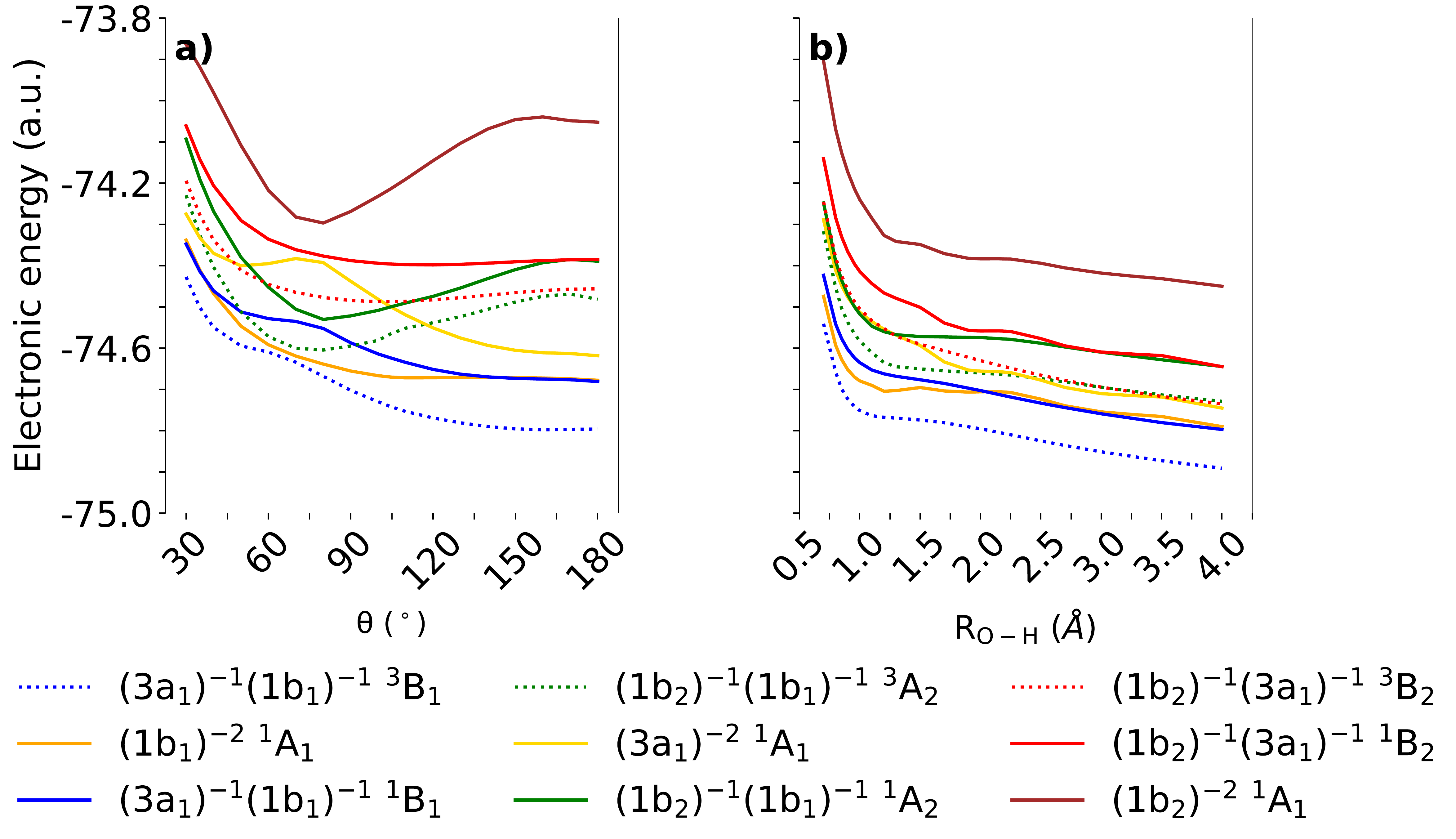}
    \caption{Potential-energy curves of the double-valence-hole states of H$_2$O  plotted as in  Figure~\ref{Fig:SVH_PECs}.}
    \label{Fig:DVH_PECs}
\end{figure}

The  PESs as a function of the bond angle and bond distance are shown in the two subplots of Figure~\ref{Fig:DVH_PECs}, with triplet and singlet states denoted as dotted and solid lines, respectively. 
%
The two lowest-lying triplets in energy, namely, (3a$_1)^{-1}$(1b$_1)^{-1}$ $^3$B$_1$ and (1b$_2)^{-1}$(1b$_1)^{-1}$ $^3$A$_2$, undergo an avoided crossing at around 70$^\circ$, while the same is true for the singlet states (3a$_1)^{-1}$(1b$_1)^{-1}$ $^1$B$_1$ and (1b$_2)^{-1}$(1b$_1)^{-1}$ $^1$A$_2$. The higher energy states  (3a$_1)^{-2}$ $^1$A$_1$ and (1b$_2)^{-2}$ $^1$A$_1$ also feature an avoided crossing  at around 70$^\circ$. 
The PECs for the (1b$_2)^{-1}$(1b$_1)^{-1}$  and (1b$_2)^{-1}$(3a$_1)^{-1}$ states, either of singlet or triplet spin symmetry, as well as the (3a$_1)^{-1}$(1b$_1)^{-1}$ $^1$B$_1$ and (1b$_1)^{-2}$ $^1$A$_1$ states  almost overlap at 30$^{\circ}$ and 180$^\circ$. Indeed, for small and 180$^{\circ}$ angles, the molecule becomes linear and  the  3a$_1$ and 1b$_{1}$ orbitals turn into a $\pi_u$ orbital and have the same energy. 
 Concerning the PECs as a function of the bond distance, we find that they all become repulsive with increasing bond distance. Also, the same states that overlap at small and 180$^{\circ}$ angles, overlap at large bond distances. Indeed, the escape of hydrogen atom results  in an O-H linear molecule with the 3a$_1$ and 1b$_{1}$ orbitals turning into $\pi_u$ orbitals. 
 In particular, 
 the PECs along the bond distance for the (1b$_1)^{-2}$ $^1$A$_1$, (3a$_1)^{-2}$ $^1$A$_1$ and (1b$_2)^{-1}$(3a$_1)^{-1}$ $^1$A$_1$ states, exhibit the same repulsive trends as those reported by Nobusada~\textit{et al.}~\cite{Nobusada:2000}, who computed the corresponding curves using the MRCI method.
\begin{table}[H]
\centering
\begin{threeparttable}
\begin{adjustbox}{max width=\textwidth, max totalheight=0.75\textheight}

\begin{tabular}{c| c| c c}
\hline
\hline
 & & \multicolumn{2}{c}{Vertical Ionization Potential (eV)}\\

Electron configuration & State & Our work & Literature \\
\hline

(3a$_1)^{-1}$ (1b$_1)^{-1}$& $^3$B$_1$ &
39.63(SA-CASSCF) & 40.33 [a]\\[6pt]

& $^1$B$_1$ &
42.84(SA-CASSCF) & 42.84 [a]\\[6pt]

(1b$_1)^{-2}$& $^1$A$_1$ &
41.57(SA-CASSCF) & 41.37 [a]\\
                &              &                 41.39(MRCI)     & 41.30 [b]\\
                &              &                 & 41.10 [c]\\
 & & &\\

(1b$_2)^{-1}$ (1b$_1)^{-1}$& $^3$A$_2$ &
44.42(SA-CASSCF) & 44.32 [a]\\[6pt]

& $^1$A$_2$ &
46.20(SA-CASSCF) & 46.02 [a]\\[6pt]\\
 (1b$_2)^{-1}$ (3a$_1)^{-1}$& $^3$B$_2$& 46.55(SA-CASSCF)&46.26 [a]\\ \\
 & $^1$B$_2$ &
49.03(SA-CASSCF) & 48.36 [a]\\
                &              &                 & 47.66 [d]\\\\

(3a$_1)^{-2}$& $^1$A$_1$ &
46.22(SA-CASSCF) & 46.03 [a]\\
                &              &                 45.97(MRCI)     & 46.30 [b]\\
                &              &                 & 45.90 [c]\\\\

(1b$_2)^{-2}$& $^1$A$_1$ &
53.98(SA-CASSCF) & 53.20 [b]\\
                &              &                 & 52.90 [c]\\
                &              &                 & 53.59 [d]\\\\

(2a$_1)^{-1}$ (1b$_1)^{-1}$& $^3$B$_1$ &
56.88(SA-TS-CASSCF) & 57.5 [c]\\
                    &              &                  & 59.16 [d]\\
                    &              &                 & 57.1 [e]\\
 & & &\\
 & $^1$B$_1$ &
65.04(SA-TS-CASSCF) & 62.57 [d]\\
                    &              &               & 63.90 [e]\\[4pt]

(2a$_1)^{-1}$ (3a$_1)^{-1}$& $^3$A$_1$ &
61.81(SA-TS-CASSCF) & 61.04 [d]\\[6pt]

& $^1$A$_1$ &
68.83(SA-TS-CASSCF)& 65.1 [c]\\
      &       &      & 63.87 [d]\\[8pt]

(2a$_1)^{-1}$ (1b$_2)^{-1}$& $^3$B$_2$ &
62.65(SA-TS-CASSCF) & 65.1 [b]\\
                &              &                 & 63.87 [d]\\[8pt]
& $^1$B$_2$ &
69.96(SA-TS-CASSCF)& 70.5 [c]\\
      &       &      & 68.80 [d]\\[8pt]

(2a$_1)^{-2}$& $^1$A$_1$ &
83.71(SA-TS-CASSCF) & 82.3 [b]\\
                 &              &                  & 83.60 [d]\\
                 &              &                 & 83.30 [e]\\[4pt]
 (1a$_1)^{-1}$(1b$_1)^{-1}$& $^3$B$_1$& 566.88(SA-TS-CASSCF)&567.46 [f]\\ \\
 & $^1$B$_1$& 570.36(SA-TS-CASSCF)&571.06 [f]\\ \\
 (1a$_1)^{-1}$(3a$_1)^{-1}$& $^3$A$_1$& 568.84(SA-TS-CASSCF)&569.49 [f]\\ \\
  & $^1$A$_1$ & 572.25(SA-TS-CASSCF)&572.77 [f]\\ \\
 (1a$_1)^{-1}$ (1b$_2)^{-1}$& $^3$B$_2$& 572.71(SA-TS-CASSCF)&573.81 [f]\\ \\
 & $^1$B$_2$ & 575.10(SA-TS-CASSCF)&576.05 [f]\\ \\
 (1a$_1)^{-1}$(2a$_1)^{-1}$& $^3$A$_1$& 589.35(SA-TS-CASSCF)&589.84 [f]\\ \\
 & $^1$A$_1$ & 594.72(SA-TS-CASSCF)&594.77 [f]\\ \\
 (1a$_1)^{-2}$& $^1$A$_1$ & 1171.41(TS-CASSCF)&1170.85 [f]\\
 & & &1171~$\pm$~1 [g]\\
 \hline
 \hline
\end{tabular}
\end{adjustbox}

\begin{tablenotes}[para]
\item \hfill\
\parbox{1.2\linewidth}{
$^{[a]}$MRCI calculations \cite{Gervais:2009} 
$^{[b]}$Auger-spectroscopy measurements \cite{Siegbahn:1975} 
$^{[c]}$Electrostatic-electron-spectrometry measurements \cite{Moddeman:1971}
$^{[d]}$Diffusion Quantum Monte Carlo (DQMC) calculations \cite{Streit:2009} 
$^{[e]}$Second-order algebraic-diagrammatic-construction (ADC(2)) calculations \cite{Tarantelli:1985}
$^{[f]}$SA-CASSCF calculations \cite{Mucke:2013}   
$^{[g]}$Double-Auger-decay spectrum \cite{Mucke:2013}.
}
\hfill
\end{tablenotes}
\end{threeparttable}

\caption{Vertical ionization potentials for the doubly charged ions of H$_2$O. The ions with two electrons missing from the 1b$_1$ (1a$''$ in C$_s$), 3a$_1$ (4a$'$ in C$_s$) or 1b$_2$ (3a$'$ in C$_s$) orbitals are DVH states; the ions with one electron missing from either the 1b$_1$, 3a$_1$ or 1b$_2$ and one electron missing from either the 2a$_1$ (2a$'$ in C$_s$) or 1a$_1$ (1a$'$ in C$_s$) orbitals are DCVH  states. The ions with two electrons  missing from 
  the 2a$_1$ (2a$'$ in C$_s$) orbital and/or the 1a$_1$ (1a$'$ in C$_s$) orbital are DCH states. }
\label{Table:DCH}
\end{table}

\begin{figure}[H]
    \centering
    \includegraphics[width=1.0\linewidth]{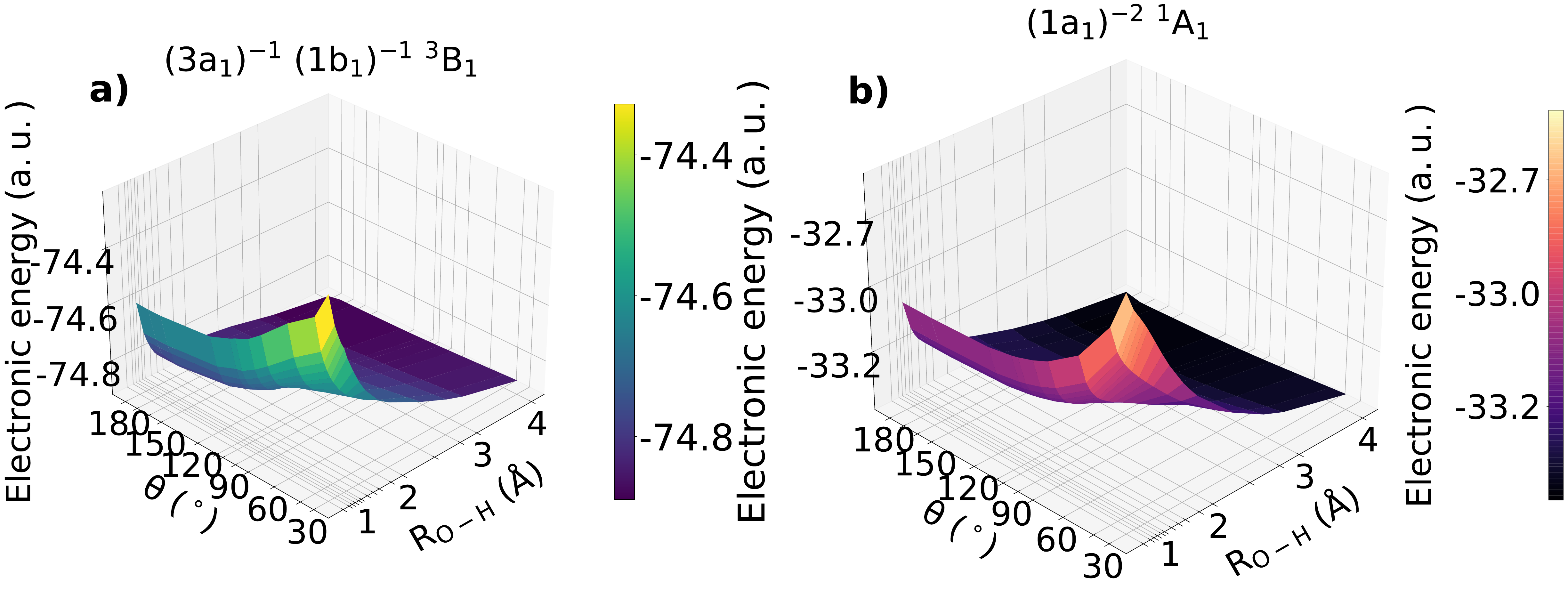}
    \caption{Potential-energy surfaces for the a) (3a$_1)^{-1}$(1b$_1)^{-1}$ $^3$B$_1$ and  b) (1a$_1)^{-2}$ $^1$A$_1$ ion states of H$_2$O as plotted in Figure~\ref{Fig:SVH-SCH_PES}.}
    \label{Fig:DVH-DCH_PES}
\end{figure}

In Figure~\ref{Fig:DVH-DCH_PES} a), we also plot the PES of (3a$_1)^{-1}$(1b$_1)^{-1}$ $^3$B$_1$ as a function of the bond angle and one bond distance, with the other distance fixed at the equilibrium distance of neutral water.  A cut through this PES at bond distance 0.956~\AA~corresponds to the PEC in Figure~\ref{Fig:DVH_PECs} a), while a cut at bond angle 104.5 $^{\circ}$ corresponds to the PEC  in Figure~\ref{Fig:DVH_PECs} b).

We now consider the ions where at least one electron is missing from the core orbitals, i.e., the DCH and DCVH  states. 
 The PESs for these ions are computed with the larger active space and the aug-cc-pCVQZ basis set. 
 The SA-TS-CASSCF technique was employed for all these ion states, except for the (1a$_1)^{-2}$ $^1$A$_1$ state where we employ   TS-CASSCF.  We use two-step (TS) optimization, with  core orbitals being optimized in a different step than valence orbitals due to the presence of at least one core hole, whereas we  state average (SA) to account for avoided crossings.  The (1a$_1)^{-2}$ $^1$A$_1$ ion state has much higher energy compared to the other DCH and DCVH ions, resulting in SA not being necessary.

The VIPs  for these ion states are listed in Table~\ref{Table:DCH}. The ions where at least one electron is missing from the 2a$_1$ orbital are benchmarked against the Diffusion Quantum Monte Carlo (DQMC) method of Streit~\textit{et al.}~\cite{Streit:2009} and the ADC(2) computations of Tarantelli~\textit{et al.}~\cite{Tarantelli:1985},  the Auger-spectroscopy measurements of Siegbahn~\textit{et al.}~\cite{Siegbahn:1975} and the electrostatic-electron spectrometry measurements of Moddeman~\textit{et al.}~\cite{Moddeman:1971}. 
The ions where at least one electron is missing from the 1a$_1$ orbital are assessed against the SA-CASSCF calculations of Mucke~\textit{et al.}~\cite{Mucke:2013}.
Finally, the VIP obtained for the (1a$_1)^{-2}$ $^1$A$_1$ state is compared with both the SA-CASSCF calculations and the Double-Auger spectroscopy measurement reported in Ref.~\cite{Mucke:2013}.
Our calculations are overall in very good agreement with the literature. In particular, our result for the VIP of the (1a$_1)^{-2}$ $^1$A$_1$ state falls within the error bars of the experiment in Ref.~\cite{Mucke:2013}. The only exception is our result of 68.83 eV  for the VIP of  the (2a$_1)^{-1}$(3a$_1)^{-1}$ $^1$A$_1$ state, which overestimates by 3.7 eV  the measurement in Ref.~\cite{Moddeman:1971}.

\begin{figure}
    \centering
    \includegraphics[width=0.9\linewidth]{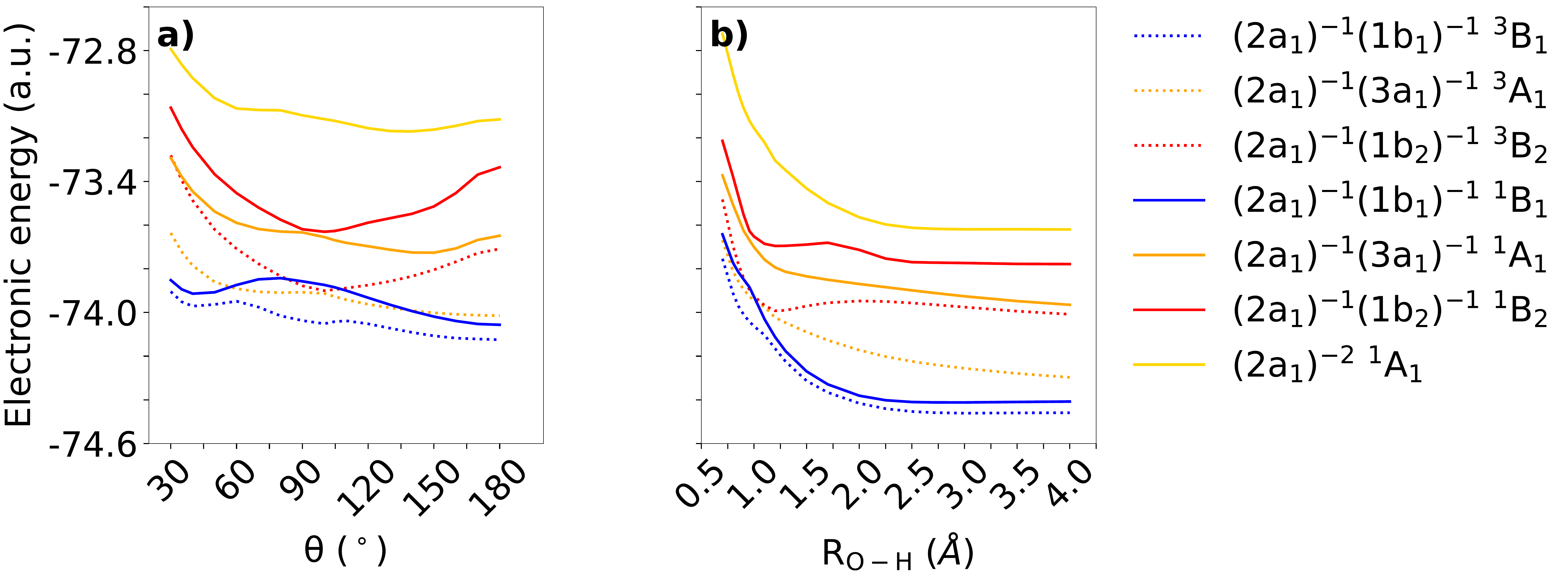}
    \caption{Potential-energy curves of the (2a$_1)^{-1}$V$^{-1}$ DCVH states and the (2a$_1)^{-2}$ DCH of H$_2$O plotted as in Figure~\ref{Fig:SVH_PECs}.}
    \label{Fig:DCH_PECs}
\end{figure}

%
We plot the PECs of the ions where at least one electron is missing from the 2a$_1$ orbital in  Figure~\ref{Fig:DCH_PECs} as a function of the bond angle a) and distance b).
The (2a$_1)^{-1}$(3a$_1)^{-1}$ $^3$A$_1$ and (2a$_1)^{-1}$(1b$_2)^{-1}$ $^3$B$_2$ triplet-spin-symmetry states and their respective singlet-spin symmetry states undergo avoided crossings in a).  The PEC of the DCH (2a$_1)^{-2}$ $^1$A$_1$ state is well separated in energy from the lower-energy DCVH states both in a) and b).  Moreover, we find that the PECs of all the DCVH and DCH states with at least one electron missing from the 2a$_{1}$ orbital are repulsive as a function of the bond distance at larger distances. Also, the states $^3$A$_1$ and $^3$B$_2$ undergo an avoided crossing as a function of the bond distance.

\begin{figure}[H]
    \centering
    \includegraphics[width=0.9\linewidth]{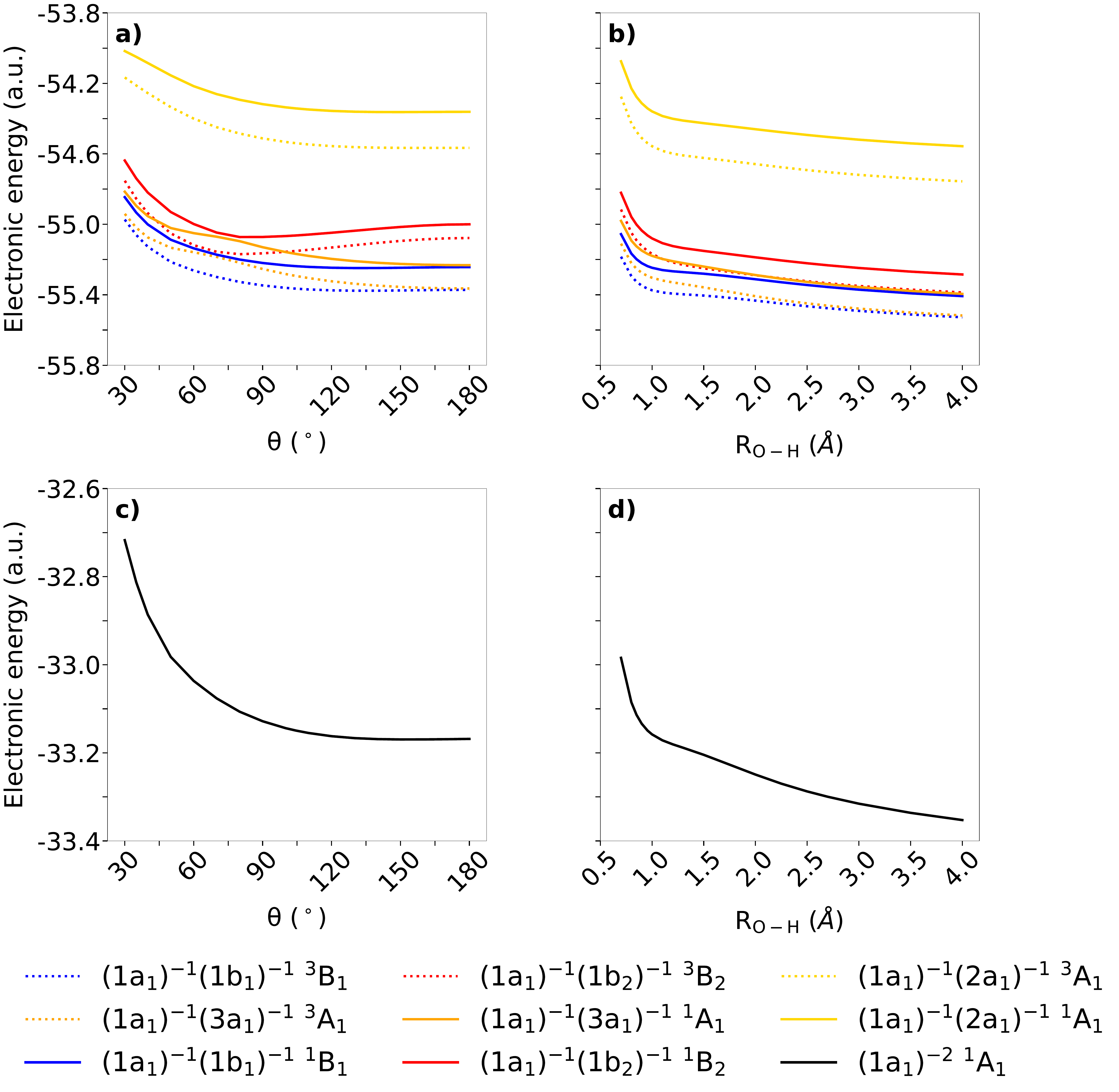}
    \caption{Potential-energy curves of the (1a$_1)^{-1}$V$^{-1}$ DCVH and the (1a$_1)^{-2}$ DCH states of H$_2$O plotted as in Figure~\ref{Fig:SVH_PECs}.}
    \label{Fig:1a1-1V-1_PECs}
\end{figure}

For the DCVH (1a$_1)^{-1}$V$^{-1}$ states and the DCH (1a$_1)^{-1}$(2a$_1)^{-1}$ ion, we plot the PECs in Figure~\ref{Fig:1a1-1V-1_PECs} a) and b). The (1a$_1)^{-1}$(3a$_1)^{-1}$ $^3$A$_1$ and the 1a$_1)^{-1}$(1b$_2)^{-1}$ $^3$B$_2$ triplet states as well as their respective singlet states undergo avoided crossings as a function of the bond angle. Also, 
 the (1a$_1)^{-1}$(1b$_1)^{-1}$ and (1a$_1)^{-1}$(3a$_1)^{-1}$ electronic-configuration triplet (singlet) states become degenerate towards 180$^\circ$, since the 3a$_1$ and 1b$_{1}$ molecular orbitals become  $\pi_{u}$ orbitals of the linear molecule.  Moreover, all PECs are repulsive as a function of the bond distance. In Figure~\ref{Fig:1a1-1V-1_PECs} c) and d), we plot separately the  (1a$_1)^{-2}$ $^1$A$_1$ state, since the other DCVH states are  significantly lower in energy.  We find that the PEC is repulsive both as a function of bond angle and distance due to the strong repulsion of the nuclei in the absence of the two inner most electrons. Also,  we plot the PES of (1a$_1)^{-2}$ $^1$A$_1$ in Figure~\ref{Fig:DVH-DCH_PES} b) as a function of one bond distance and bond angle. 

\subsection{Triply charged ion states}

%
First, we consider the triply-valence-hole (TVH) states, which are the triply charged water ions lowest in energy.
 We compute the PESs of these ions with the larger active space and the aug-cc-pVQZ basis set, employing the SA-CASSCF technique (see Table~\ref{Table:Approach}), since many TVH states are close in energy and undergo avoiding crossings.  States with electronic configurations with three electrons missing from three different valence orbitals have total spin equal to either 1/2 (\textit{doublet} spin symmetry) or  3/2 (\textit{quartet} spin symmetry)~\cite{PopleBOOK}.
 
 In Table~\ref{Table:TVH}, we compare  our results for the VIPs of the TVH states  with the SA-CASSCF calculations of Mucke~\textit{et al.}~\cite{Mucke:2013}, and the MRCI and ADC(2) computations of Handke~\textit{et al.}~\cite{Handke:1995}. We find that the VIPs range from 83 eV to 95 eV and are larger than the values corresponding to the SA-CASSCF benchmark \cite{Mucke:2013} and lower than the ones obtained with ADC(2) \cite{Handke:1995}.
The only exception is the (1b$_2)^{-2}$(1b$_1)^{-1}$ $^2$B$_1$ state, which is 0.21 eV lower than the SA-CASCF calculations in Ref.~\cite{Mucke:2013} 
Overall, our results for the VIPs have very good agreement with the results presented in  Refs \cite{Mucke:2013,Handke:1995}.

\begin{table}[H]
\centering
\begin{threeparttable}
\begin{adjustbox}{max width=\textwidth, max totalheight=0.9\textheight}

\begin{tabular}{c|c|cc}
\hline
\hline
\multirow{2}{*}{Electron configuration}                                 & \multirow{2}{*}{State} & Vertical Ionization Potential (eV)&                                                                                                         \\
                                                                        &                        & Ours    & Literature     \\
                                                                        \hline
(3a$_1)^{-1}$(1b$_1)^{-2}$& $^2$A$_1$& 83.73   & 82.82 [a]\\
 & & &84.99 [b]\\
 & & &\\
(3a$_1)^{-1}$(1b$_2)^{-1}$(1b$_1)^{-1}$& $^4$A$_2$& 84.02   & 83.53 [a]\\
 & & &85.50 [b]\\
 & & &\\
 & $^2$A$_2$& 87.50   &87.34 [a]\\
 & & &88.99 [b]\\
 & & &\\
 & $^2$A$_2$& 88.34   &88.10 [a]\\
 & & &89.64 [b]\\
 & & &\\
(3a$_1)^{-2}$(1b$_1)^{-1}$& $^2$B$_1$& 85.70   & 84.71 [a]\\
 & & &86.87 [b]\\
 & & &\\
(1b$_2)^{-1}$ (1b$_1)^{-2}$& $^2$B$_2$& 87.01   & 86.95 [a]\\
 & & &88.49 [b]\\
 & & &\\
(1b$_2)^{-1}$ (3a$_1)^{-2}$& $^2$B$_2$& 91.79   & 91.18 [a]\\
 & & &92.88 [b]\\
 & & &\\
(1b$_2)^{-2}$ (1b$_1)^{-1}$& $^2$B$_1$& 92.54   & 92.75 [a]\\
 & & &93.96 [b]\\
 & & &\\
(1b$_2)^{-2}$ (3a$_1)^{-1}$& $^2$A$_1$& 94.38   & 94.08 [a]\\
 & & &95.49 [b]\\
 & & &\\
(2a$_1)^{-1}$ (3a$_1)^{-1}$ (1b$_1)^{-1}$& $^4$B$_1$& 99.17& 96.76 [a]\\
 & & &99.09 [b]\\
 & & &\\
 & $^2$B$_1$& 104.13&104.10 [a]\\
 & & &105.31 [b]\\
 & & &\\
 & $^2$B$_1$& 109.87&108.95 [a]\\
 & & &109.92 [b]\\
 & & &\\
(2a$_1)^{-1}$ (1b$_2)^{-1}$ (1b$_1)^{-1}$& $^4$A$_2$& 100.22& 101.66 [a]\\
 & & &103.32 [b]\\
 & & &\\
 & $^2$A$_2$& 106.94&108.28 [a]\\
 & & &109.14 [b]\\
 & & &\\
 & $^2$A$_2$& 113.42&114.17 [a]\\
 & & &114.98 [b]\\
 & & &\\
(2a$_1)^{-1}$ (1b$_1)^{-2}$& $^2$A$_1$& 102.63  & 103.08 [a]\\
 & & &104.41 [b]\\
 & & &\\
 & & &\\
 (2a$_1)^{-1}$ (1b$_2)^{-1}$ (3a$_1)^{-1}$& $^4$B$_2$& 104.95  &102.76 [a]\\
 & & &104.65 [b]\\
 & & &\\
 & $^2$B$_2$& 109.49  &109.10 [a]\\
 & & &110.17 [b]\\
 & & &\\
 & $^2$B$_2$& 115.41  &116.64 [a]\\
 & & &115.65 [c]\\
 & & &\\
(2a$_1)^{-1}$ (3a$_1)^{-2}$& $^2$A$_1$& 108.94  & 107.01 [a]\\
 & & &108.45 [b]\\
 & & &\\
(2a$_1)^{-1}$ (1b$_2)^{-2}$& $^2$A$_1$& 113.89  & 116.00 [a]\\
 & & &117.27 [b]\\
 & & &\\
(2a$_1)^{-2}$ (1b$_1)^{-1}$& $^2$B$_1$& 125.54  & 127.16 [b]\\
 & & &127.53 [c]\\
 & & &\\
(2a$_1)^{-2}$ (3a$_1)^{-1}$& $^2$A$_1$& 129.11  & 130.04 [a]\\
 & & &128.25 [c]\\
 & & &127.93 [b]\\ \\
(2a$_1)^{-2}$ (1b$_2)^{-1}$& $^2$B$_2$& 131.18  & 132.83 [a]\\
\hline
\hline
\end{tabular}
\end{adjustbox}

\begin{tablenotes}[para]
\item \hfill\
\parbox{1.2\linewidth}{
 \item $^{[a]}$SA-CASSCF calculations \cite{Mucke:2013}  
 \item $^{[b]}$MRCI calculations \cite{Handke:1995}
 \item $^{[c]}$ADC(2) calculations \cite{Handke:1995}
}
\hfill

\end{tablenotes}

\end{threeparttable}

\caption{Vertical ionization potentials (VIPs) for the triply charged states of H$_2$O with electrons missing either from valence orbitals (TVH states) or from the core 2a$_{1}$ orbital and valence orbitals (TCVH states with at least one hole in 2a$_{1}$).}
\label{Table:TVH}
\end{table}
%

\begin{figure}[H]
    \centering
    \includegraphics[width=0.9\linewidth]{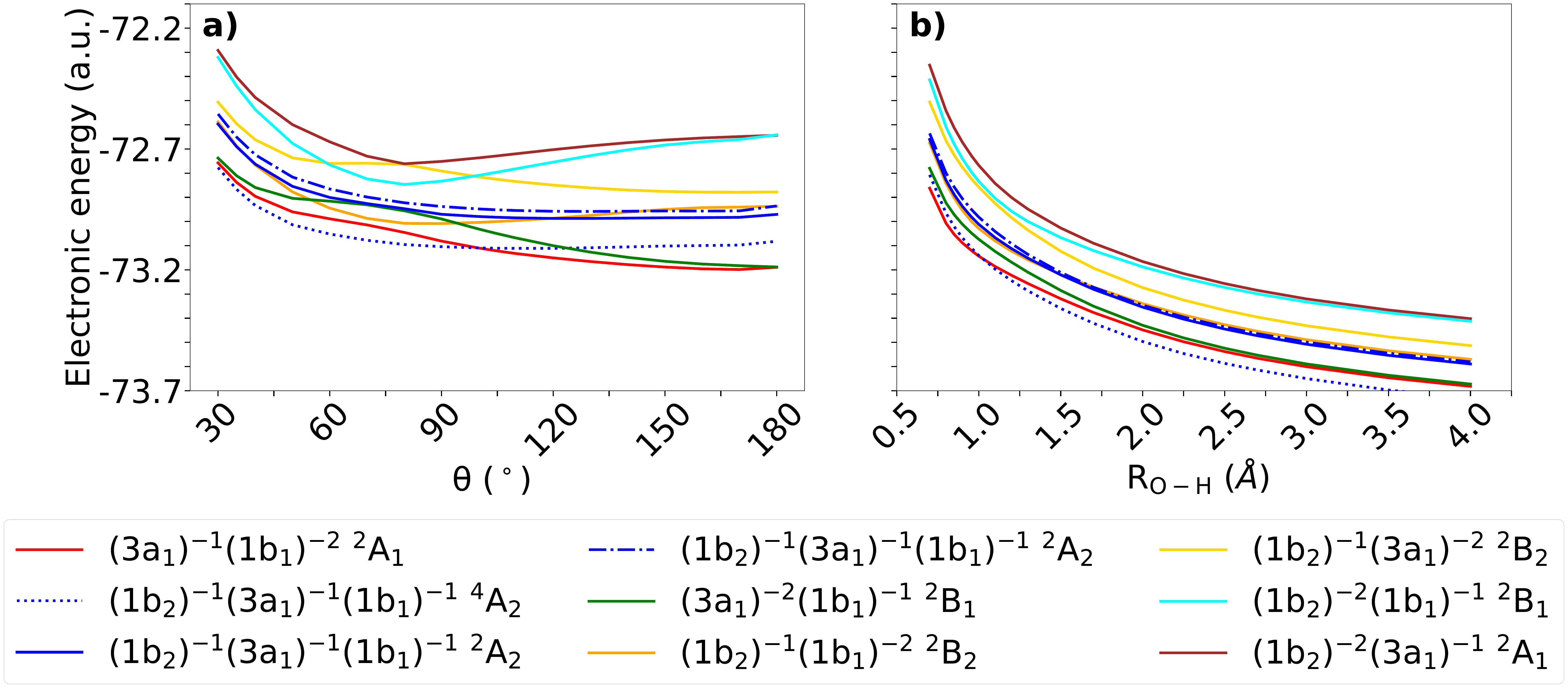}
    \caption{Potential-energy curves of the triply-valence-hole states of H$_2$O plotted as in Figure~\ref{Fig:SVH_PECs}.}
    \label{Fig:TVH_PECs}
\end{figure}

Next, we plot the PECs of the TVH states in Figure~\ref{Fig:TVH_PECs} as a function of the bond angle a) and the bond distance b). As a function of the bond angle, the states 
of the same symmetry (3a$_1)^{-1}$(1b$_1)^{-2}$ $^2$A$_1$ and (1b$_2)^{-1}$ (1b$_1)^{-2}$ $^2$B$_2$ as well as the states (3a$_1)^{-2}$(1b$_1)^{-1}$ $^2$B$_1$ and (1b$_2)^{-2}$(1b$_1)^{-1}$ $^2$B$_1$ and the states  (1b$_2)^{-1}$(3a$_1)^{-2}$ $^2$B$_2$ and (1b$_2)^{-2}$(3a$_1)^{-1}$ $^2$A$_1$ undergo avoided crossings. As a function of the bond  distance, all PECs are repulsive. Also, the two doublet states  corresponding to the (1b$_2)^{-1}$(3a$_1)^{-1}$(1b$_1)^{-1}$ electronic configuration, are very similar as a function of both angle and distance, with the lowest (highest) in energy state being the one where the close-in-energy orbitals 3$\mathrm{a_{1}}$ and 1b$_{1}$ have parallel (antiparallel) spin.

\begin{figure}[H]
    \centering
    \includegraphics[width=1.0\linewidth]{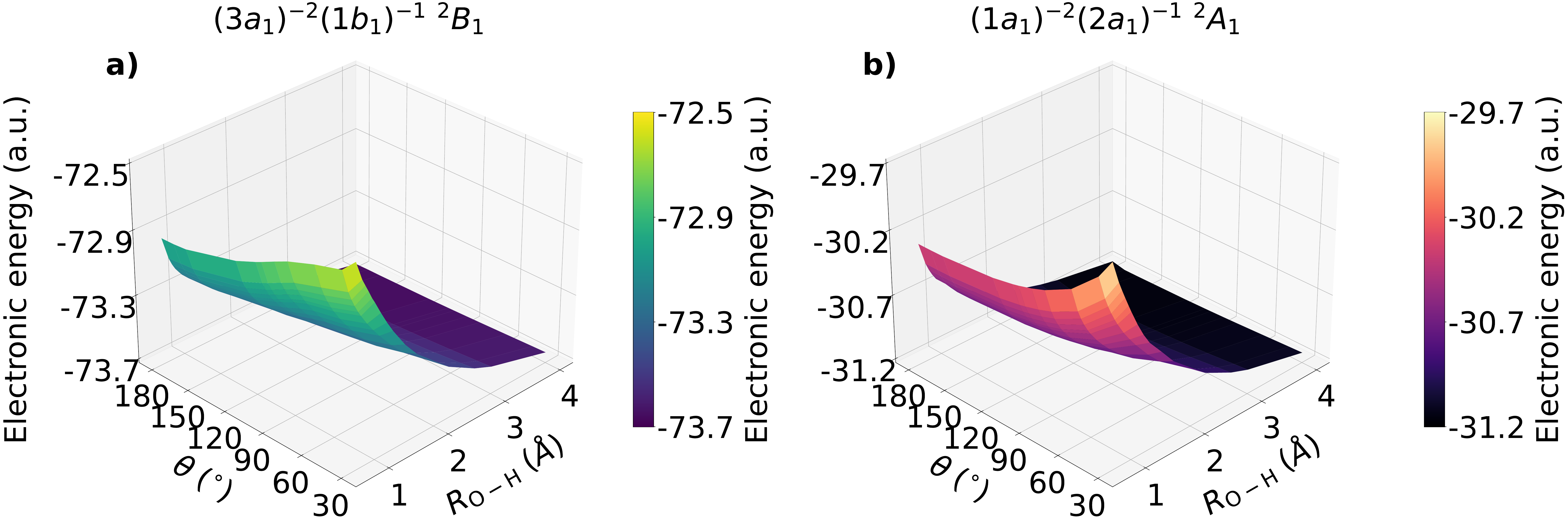}
    \caption{Potential-energy surfaces obtained with the CASSCF technique for the a) (3a$_1)^{-2}$(1b$_1)^{-1}$ $^2$B$_1$ and  b) (1a$_1)^{-2}$(2a$_1)^{-1}$ $^2$A$_1$ ion states of H$_2$O as plotted in Figure~\ref{Fig:SVH-SCH_PES}.}
    \label{Fig:TVH-TCH_PES}
\end{figure}

For the (3a$_1)^{-2}$(1b$_1)^{-1}$ $^2$B$_1$ state, we plot the PES as a function of one bond distance and the bond angle in Figure~\ref{Fig:TVH-TCH_PES} a), with the other bond distance fixed at the equilibrium bond distance of neutral water. Along the bond angle, the PES features the bump due to the avoided crossing with state (1b$_2)^{-2}$(1b$_1)^{-1}$ $^2$B$_1$ (see Figure~\ref{Fig:TVH_PECs} a)), while the behavior along the bond distance is highly  repulsive (see Figure~\ref{Fig:TVH_PECs} b)).

%
We now consider the PESs for the TCVH and TCH ion states of water, which are  calculated with the larger active space and the aug-cc-pCVQZ basis set, employing  the SA-TS-CASSCF technique, see Table~\ref{Table:Approach}. 
 We first consider the states where the core electron(s) missing are  from the 2a$_1$ orbital. The VIPs for these ions are listed in Table~\ref{Table:TVH} and benchmarked against the calculations of Mucke~\textit{et al.}~\cite{Mucke:2013} and Handke~\textit{et al.}~\cite{Handke:1995}. 
The lowest-energy TCVH ions are those where only one electron is ejected from the core orbital, with VIPs  between 99 eV and 114 eV. 
For  the (2a$_1)^{-1}$(1b$_1)^{-2}$ $^2$A$_1$ state, our computations as well as the MRCI results of Handke~\textit{et al.}\cite{Handke:1995} show that the VIP of this  state should be  lower than the VIP of the  (2a$_1)^{-1}$(1b$_2)^{-1}$(3a$_1)^{-1}$ $^4$B$_2$ state, while  the SA-CASSCF calculations of Mucke~\textit{et al.}\cite{Mucke:2013} reverse this order. We attribute this discrepancy to the smaller size of the active space and the different technique used in Ref.\cite{Mucke:2013}. 
A more substantial difference involves the VIP of the (2a$_1)^{-1}$(1b$_2)^{-2}$ $^2$A$_1$ state, with  our result of   113.89 eV  being   roughly 3 eV below the calculations of Handke~\textit{et al.} \cite{Handke:1995}.
We attribute this disagreement to the intrinsic limits of the SA-TS-CASSCF technique with respect to MRCI.  The VIPs for the other TCVH and TCH states are overall in very good agreement with Refs \cite{Mucke:2013,Handke:1995}. In  Table~\ref{Table:TVH}, we also list the VIPs of the TCVH states where two electrons are missing from  2a$_1$. These VIPs are between 125 eV and 131 eV and the difference with  \cite{Mucke:2013,Handke:1995}. is less than 2 eV.

\begin{figure}[H]
    \centering
    \includegraphics[width=0.9\linewidth]{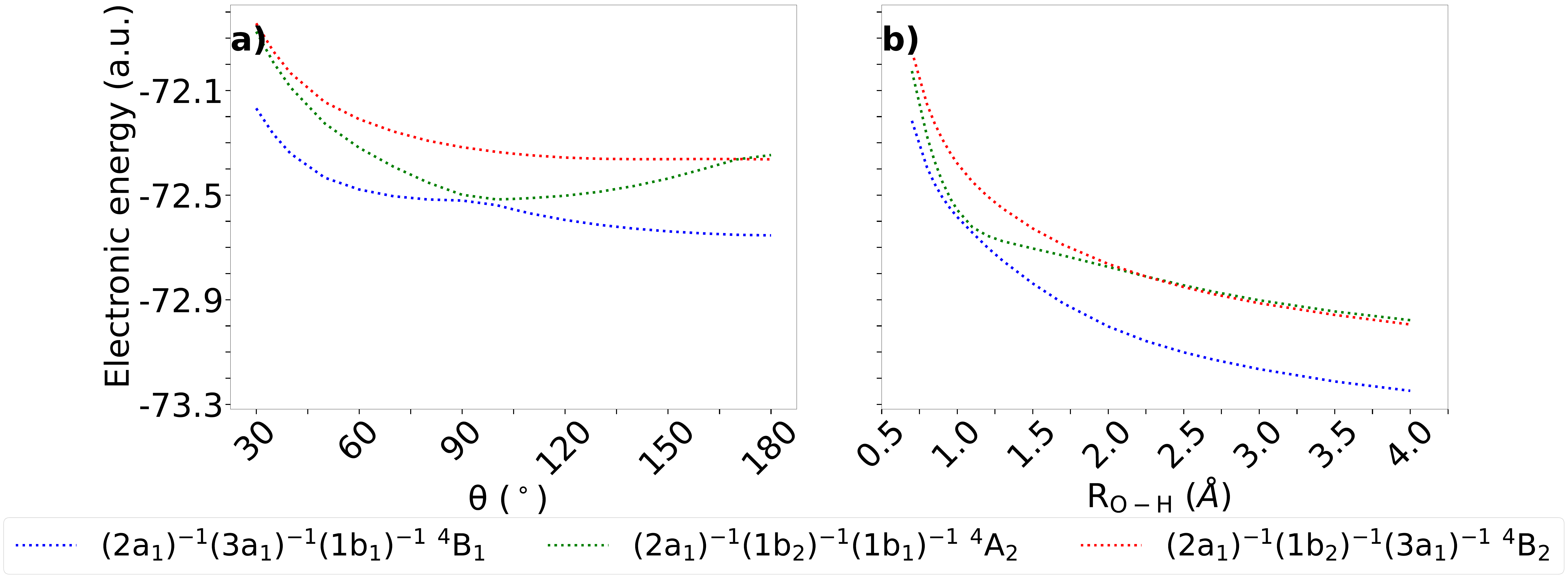}
    \caption{Potential-energy curves of the quartet TCVH states of H$_2$O with one electron missing from the 2a$_{1}$ orbital plotted as in Figure~\ref{Fig:SVH_PECs}.}
    \label{Fig:2a1-1V-2_PECs_Q}
\end{figure}

\begin{figure}[H]
    \centering
    \includegraphics[width=0.9\linewidth]{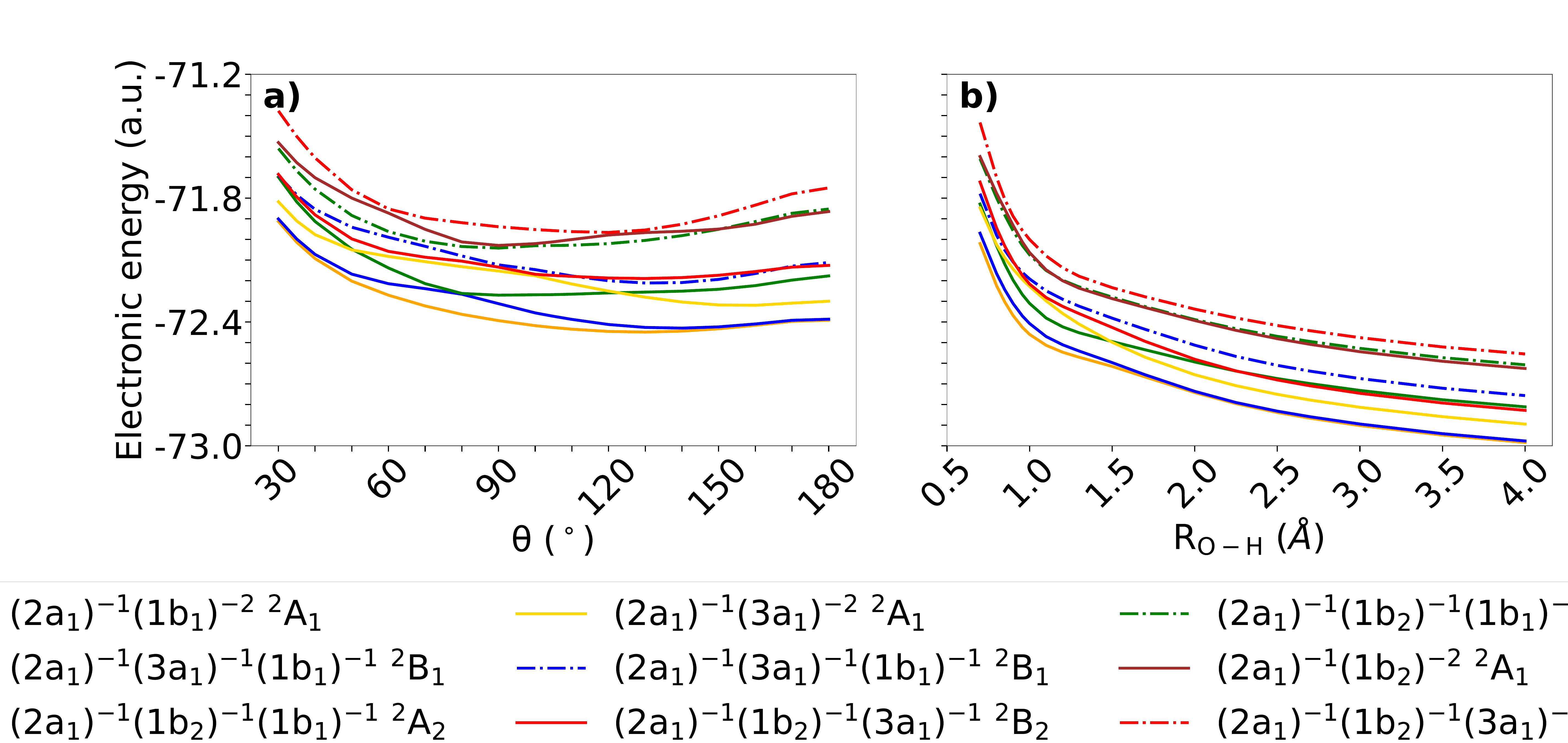}
    \caption{Potential-energy curves of the TCVH states of H$_2$O with one electron missing from the (2a$_1)^{-1}$ orbital plotted as in Figure~\ref{Fig:SVH_PECs}. }
    \label{Fig:2a1-1V-2_PECs_D}
\end{figure}

\begin{figure}[H]
    \centering
    \includegraphics[width=0.9\linewidth]{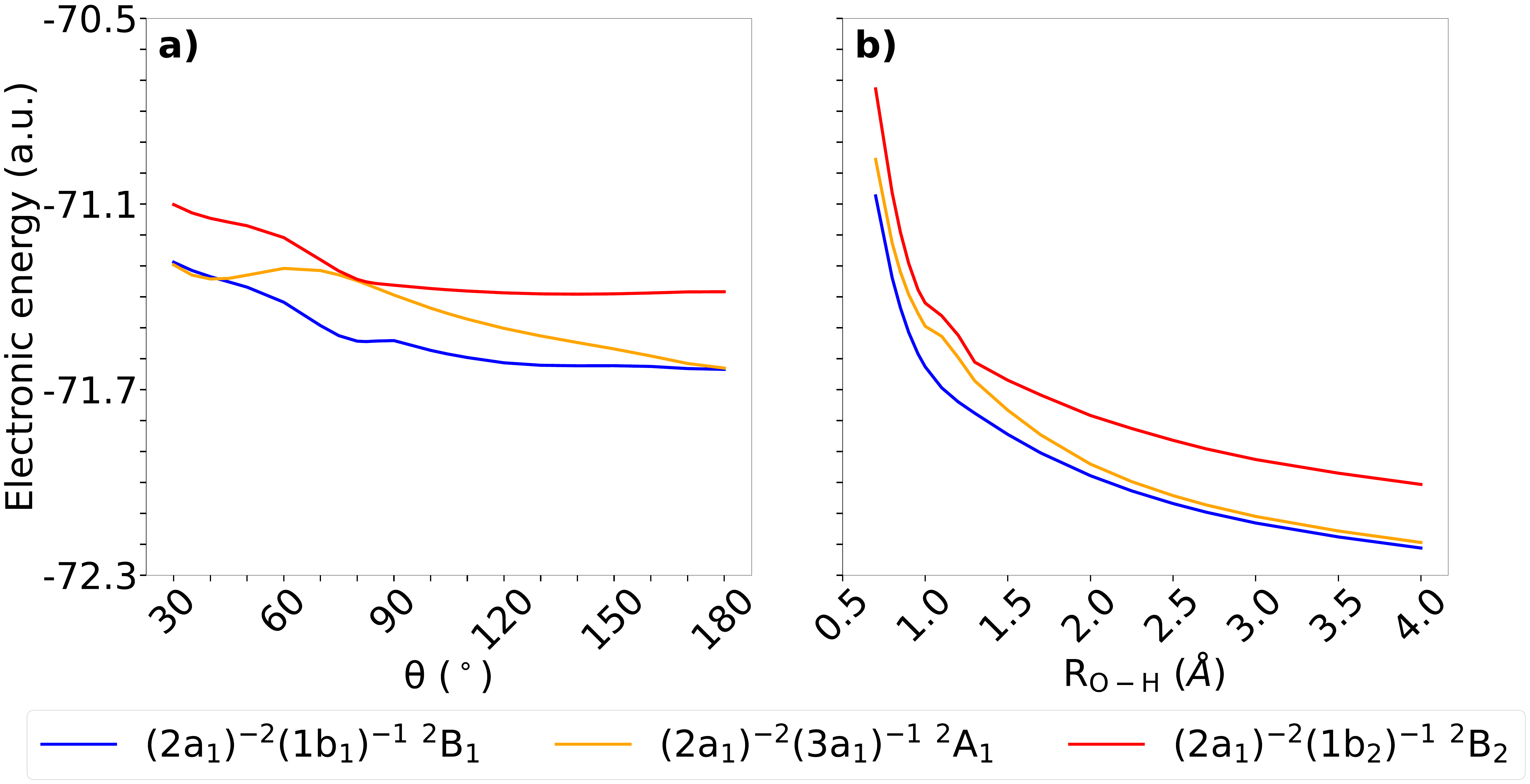}
    \caption{Potential-energy curves of the TCVH states of H$_2$O with two electrons missing from the  (2a$_1)^{-2}$ orbital plotted as in Figure~\ref{Fig:SVH_PECs}. }
    \label{Fig:2a1-2V-1_PECs}
\end{figure}

Next, we plot  the PECs of the TCVH states with one electron missing from the 2a$_1$ orbital as a function of the bond angle a) and bond distance b). The states with quartet spin symmetry  are  plotted in Figure~\ref{Fig:2a1-1V-2_PECs_Q}, while the ones with doublet spin symmetry are plotted in Figure~\ref{Fig:2a1-1V-2_PECs_D}. The quartet spin symmetry states 
(2a$_1)^{-1}$(3a$_1)^{-1}$(1b$_1)^{-1}$ $^4$B$_1$ and (2a$_1)^{-1}$(1b$_2)^{-1}$(1b$_1)^{-1}$ $^4$A$_2$ have an avoided crossing as a function of the bond angle while 
all states are repulsive as a function of the bond distance. For the doublet spin symmetry states, Figure~\ref{Fig:2a1-1V-2_PECs_D} a) shows that the same symmetry states (2a$_1)^{-1}$(3a$_1)^{-1}$(1b$_1)^{-1}$ $^2$B$_1$ and (2a$_1)^{-1}$(1b$_2)^{-1}$(1b$_1)^{-1}$ $^2$A$_2$ as well as the states (2a$_1)^{-1}$(1b$_2)^{-1}$(3a$_1)^{-1}$ $^2$B$_2$ and (2a$_1)^{-1}$(1b$_2)^{-2}$ $^2$A$_1$ undergo avoided crossings. Also, the states (2a$_1)^{-1}$(1b$_1)^{-2}$ $^2$A$_1$ and (2a$_1)^{-1}$(3a$_1)^{-1}$(1b$_1)^{-1}$ $^2$B$_1$ are degenerate for small and large angles, due to the degeneracy of the 3a$_{1}$ and 1b$_{1}$ orbitals for linear molecules, discussed previously. In Figure~\ref{Fig:2a1-1V-2_PECs_D} b), we show that all PECs of these TCVH states are repulsive.  The PECs for the TCVH ions where two electrons are missing from the 2a$_1$ orbitals are plotted in Figure~\ref{Fig:2a1-2V-1_PECs} a) and b).  The PECs are repulsive as a function of the bond distance, while as a function of the bond angle the $^2$A$_1$ and $^2$B$_2$ states undergo a strong avoided crossing.

Finally, we focus on the  core-hole states of water involving at least  an electron missing from the 1a$_{1}$ orbital. These states include TCVH states with two electrons missing from  two valence orbitals,   the 2a$_{1}$ and a valence orbital,  the 1a$_{1}$ and  a valence orbital and  TCH states where all three electrons are missing from the 1a$_{1}$ and 2a$_{1}$ core orbitals.
 The VIPs of these ions are listed in Table~\ref{Table:TCVH}. As far as we know, no other studies are available to compare.
\begin{table}[H]
\centering
\begin{tabular}{c |c |c}
\hline
\hline
\multirow{2}{*}{Electron configuration } & \multirow{2}{*}{State} &Vertical Ionization Potential (eV) \\
 &  & Ours \\
 \hline
(1a$_1)^{-1}$(3a$_1)^{-1}$(1b$_1)^{-1}$& $^4$B$_1$& 614.21 \\
 & $^2$B$_1$&619.55 \\
 & $^2$B$_1$&619.82 \\
 & &\\
(1a$_1)^{-1}$(1b$_2)^{-1}$(1b$_1)^{-1}$& $^4$A$_2$& 618.17 \\
 & $^2$A$_2$&622.27 \\
 & $^2$A$_2$&623.24 \\
 & &\\
(1a$_1)^{-1}$(1b$_1)^{-2}$& $^2$A$_1$& 618.50 \\
 & &\\
 (1a$_1)^{-1}$(1b$_2)^{-1}$(3a$_1)^{-1}$& $^4$B$_2$&619.30 \\
 & $^2$B$_2$&623.63 \\
 & $^2$B$_2$&624.34 \\
 & &\\
(1a$_1)^{-1}$(3a$_1)^{-2}$& $^2$A$_1$& 622.10 \\
 & &\\
(1a$_1)^{-1}$(1b$_2)^{-2}$& $^2$A$_1$& 627.87 \\
 & &\\
(1a$_1)^{-1}$(2a$_1)^{-1}$(1b$_1)^{-1}$& $^4$B$_1$& 632.83 \\
 & $^2$B$_1$&639.50 \\
 & $^2$B$_1$&643.91 \\
 & &\\
(1a$_1)^{-1}$(2a$_1)^{-1}$(3a$_1)^{-1}$& $^4$A$_1$& 638.12 \\
 & $^2$A$_1$&640.60 \\
 & $^2$A$_1$&647.33 \\
 & &\\
(1a$_1)^{-1}$(2a$_1)^{-1}$(1b$_2)^{-1}$& $^4$B$_2$& 638.56 \\
& $^2$B$_2$& 640.85 \\
& $^2$B$_2$& 649.24 \\
 & &\\
(1a$_1)^{-1}$(2a$_1)^{-2}$& $^2$A$_1$& 655.88 \\
 & &\\
(1a$_1)^{-2}$(1b$_1)^{-1}$& $^2$B$_1$& 1223.35 \\
 & &\\
(1a$_1)^{-2}$(3a$_1)^{-1}$& $^2$A$_1$& 1224.08 \\
 & &\\
(1a$_1)^{-2}$(1b$_2)^{-1}$& $^2$B$_2$& 1227.04 \\
 & &\\
(1a$_1)^{-2}$(2a$_1)^{-1}$& $^2$A$_1$& 1242.17 \\
\hline
\hline
\end{tabular}
\caption{Vertical ionization potentials (VIPs) for the triply charged states of H$_2$O with at least one electron missing from the core 1a$_{1}$ orbital.}
\label{Table:TCVH}
\end{table}

\begin{figure}[H]
    \centering
    \includegraphics[width=0.9\linewidth]{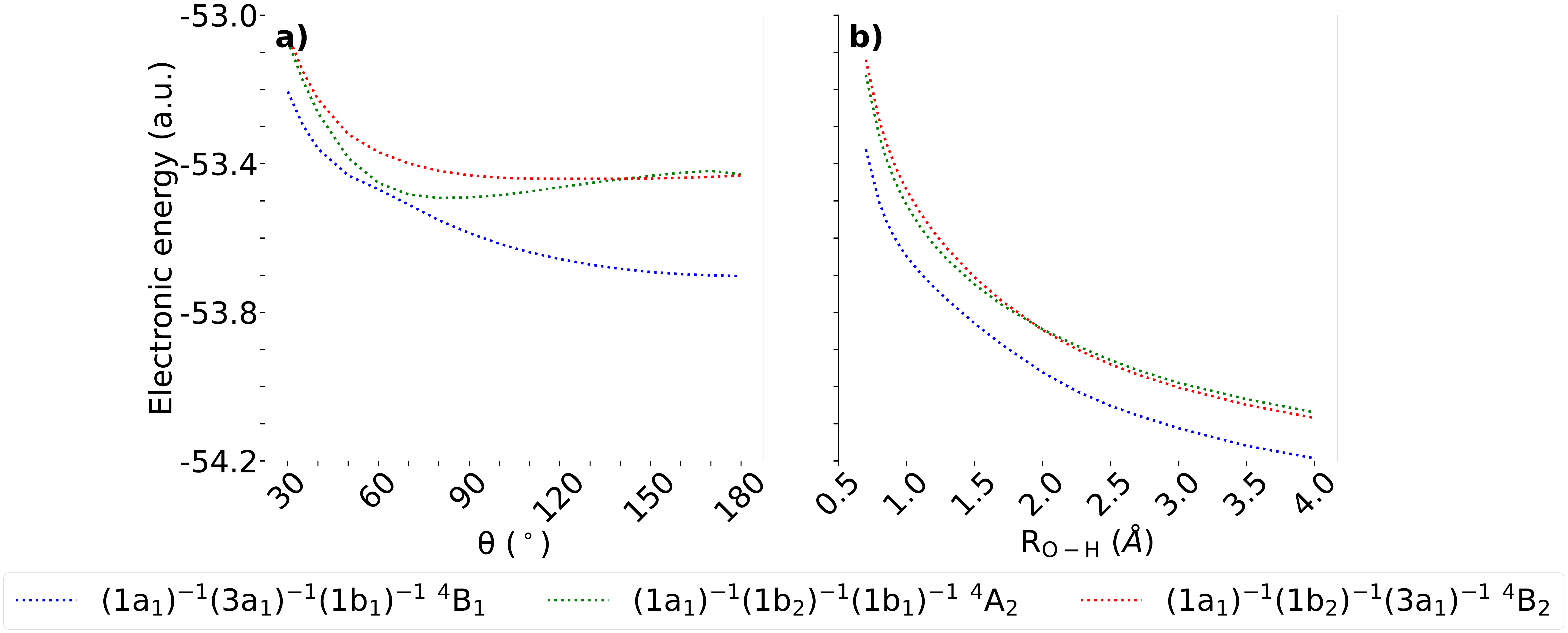}
    \caption{Potential-energy curves of the quartet TCVH states  of H$_2$O with an electron missing from the  1a$_1$ orbital and two electrons from valence orbitals plotted as in  Figure~\ref{Fig:SVH_PECs}.}
        \label{Fig:1a1-1V-2_PECs_Q}
\end{figure}

\begin{figure}[H]
    \centering
    \includegraphics[width=0.9\linewidth]{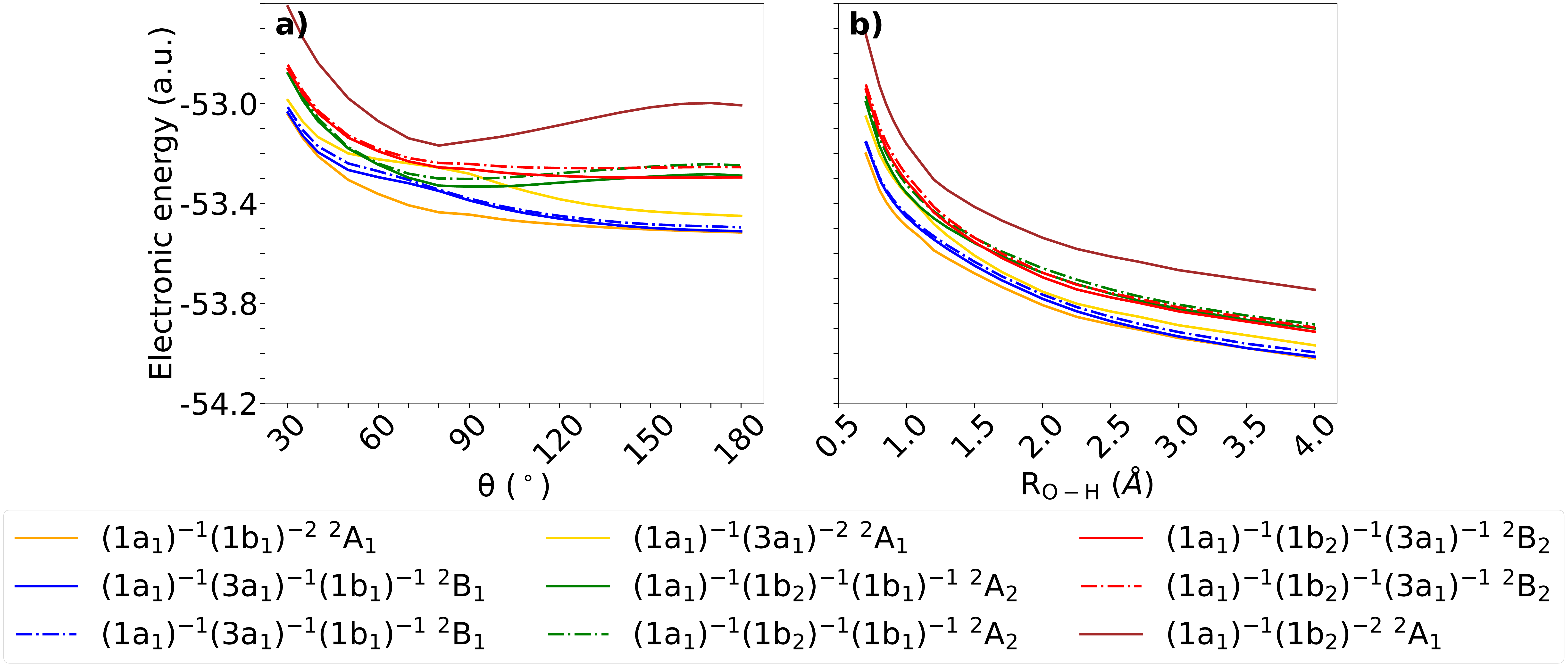}
    \caption{Potential-energy curves of the doublet TCVH states  of H$_2$O with an electron missing from the  1a$_1$ orbital and two electrons from valence orbitals plotted as in  Figure~\ref{Fig:SVH_PECs}.}
    \label{Fig:1a1-1V-2_PECs_D}
\end{figure}

Next, we plot the PECs for the TCVH states with one electron missing from the 1a$_{1}$ orbital and two electrons missing from two valence orbitals. In Figure~\ref{Fig:1a1-1V-2_PECs_Q} a) and b), we plot the quartet states. We find an avoided crossing between the same symmetry states (1a$_1)^{-1}$(3a$_1)^{-1}$(1b$_1)^{-1}$ $^4$B$_1$ and (1a$_1)^{-1}$(1b$_2)^{-1}$(1b$_1)^{-1}$ $^4$A$_2$ as a function of the bond angle, while all three PECs are repulsive as a function of the bond distance.  In Figure~\ref{Fig:1a1-1V-2_PECs_D} a) and b), we plot the doublet TCVH states. As a function of the bond angle, the states (1a$_1)^{-1}$(3a$_1)^{-1}$(1b$_1)^{-1}$ $^2$B$_{1}$ and (1a$_1)^{-1}$(1b$_2)^{-1}$(1b$_1)^{-1}$ $^2$A$_{2}$ as well as the states (1a$_1)^{-1}$(3a$_1)^{-2}$ $^2$A$_1$ and (1a$_1)^{-1}$(1b$_2)^{-1}$(3a$_1)^{-1}$ $^2$B$_2$  undergo avoided crossings. Also, the states  (1a$_1)^{-1}$(1b$_1)^{-2}$ $^2$A$_1$ and the doublet states (1a$_1)^{-1}$(3a$_1)^{-1}$(1b$_1)^{-1}$ $^2$B$_1$ are degenerate for small and large angles, due to the degeneracy of the 3a$_{1}$ and 1b$_{1}$ orbitals for linear molecules, discussed previously. As a function of bond distance all PECs are repulsive.

\begin{figure}[H]
    \centering
    \includegraphics[width=0.9\linewidth]{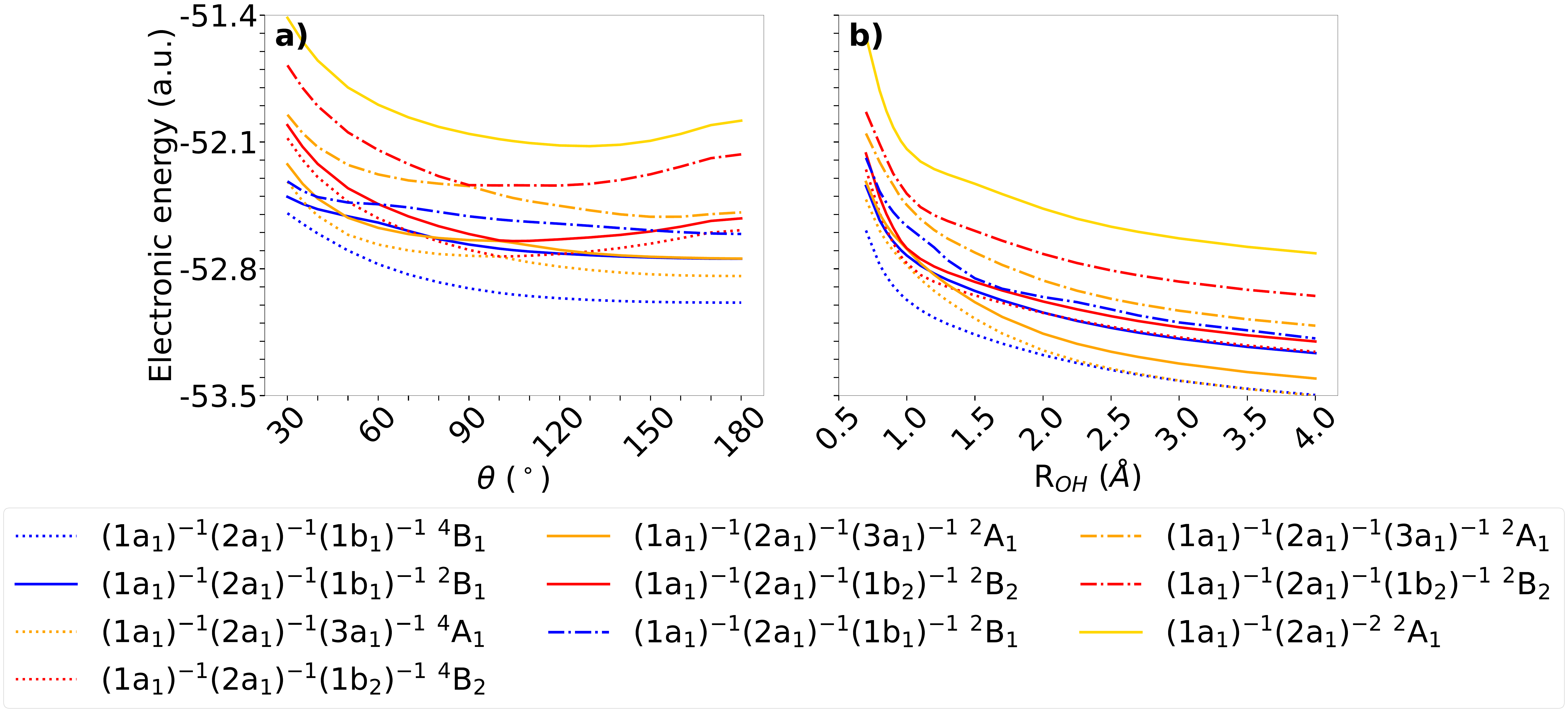}
    \caption{Potential-energy curves of the TCVH  states of H$_2$O with one electron missing from the 1a$_{1}$ orbital and at least one electron missing from the 2a$_{1}$ orbital plotted as in Figure~\ref{Fig:SVH_PECs}.}
    \label{Fig:1a1-12a1-1V-1_PECs}
\end{figure}

\begin{figure}[H]
    \centering
    \includegraphics[width=0.9\linewidth]{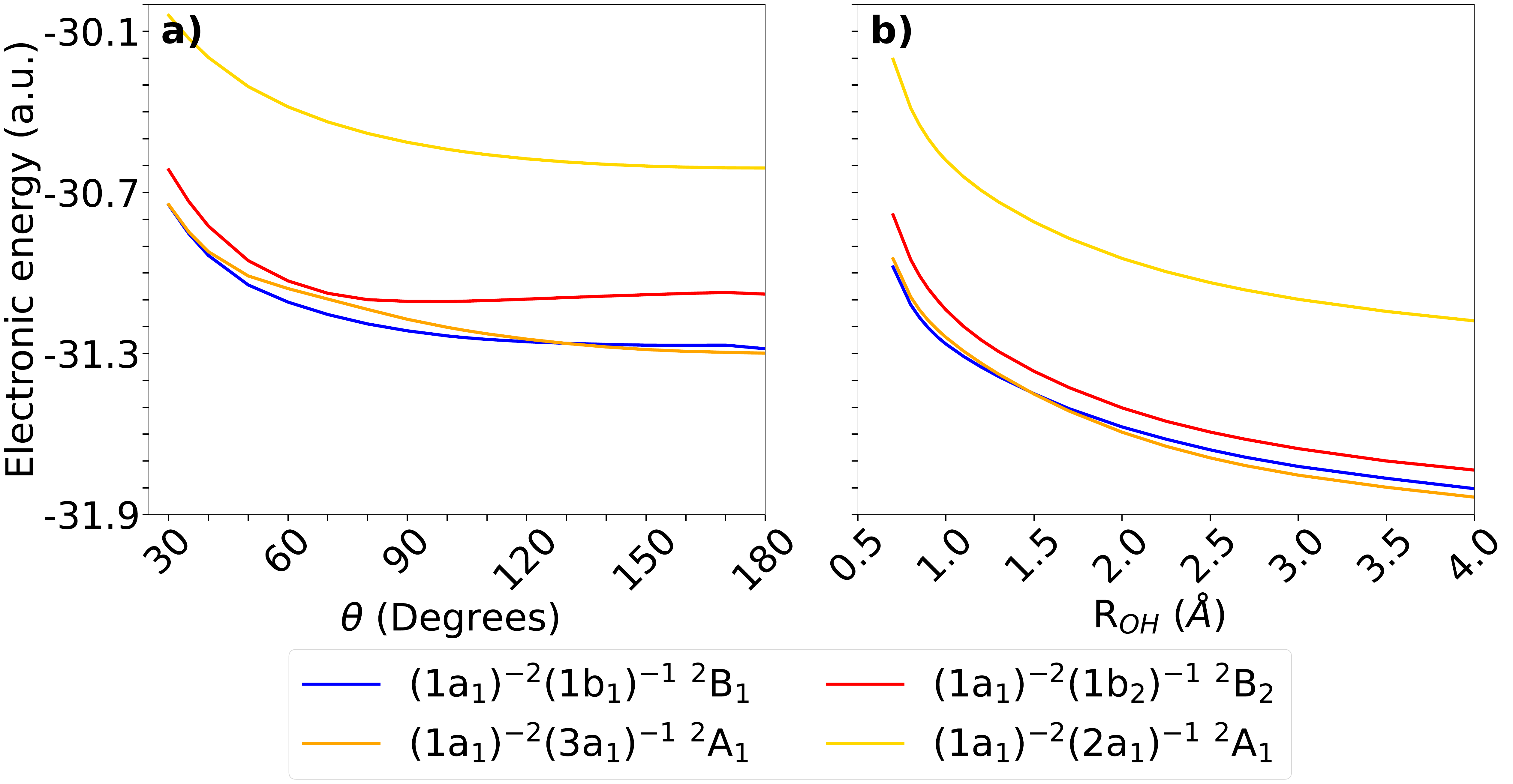}
    \caption{Potential-energy curves of the TCVH and TCH states with two electrons missing from the core 1a$_{1}$ orbital plotted as in Figure~\ref{Fig:SVH_PECs}.}
   \label{Fig:1a1-2V-1_PECs}
\end{figure}

In Figure~\ref{Fig:1a1-12a1-1V-1_PECs} a) and b), we plot the PECs for the TCVH states with one  electron missing from the 1a$_1$ and at least one electron missing from the 2a$_1$ orbital. We find that the same symmetry states (1a$_1)^{-1}$(2a$_1)^{-1}$(3a$_1)^{-1}$ $^4$A$_1$ and (1a$_1)^{-1}$(2a$_1)^{-1}$(1b$_2)^{-1}$ $^4$B$_2$, as well as the same electronic configuration states  with doublet spin symmetry undergo avoided crossings as a function of the bond angle. All these PECs are repulsive as a function of the bond distance. In Figure~\ref{Fig:1a1-2V-1_PECs} a) and b), we plot  
 the PECs for the TCVH and TCH states with  2 electrons missing from the 1a$_{1}$ orbital. We find that these PECs are repulsive as a function of both the bond angle and distance, except for the (1a$_1)^{-2}$(1b$_2)^{-1}$ $^2$B$_2$ along the bond angle, while the  states (1a$_1)^{-2}$(3a$_1)^{-1}$ $^2$A$_1$ and (1a$_1)^{-2}$(1b$_2)^{-1}$ $^2$B$_2$ undergo an avoided crossing.
  Finally, we  plot the PES for the (1a$_1)^{-2}$(2a$_1)^{-1}$ $^2$A$_1$ state as a function of  the bond angle and distance in Figure~\ref{Fig:TVH-TCH_PES} b). The PES is highly repulsive along both degrees of freedom, as is the case for the cuts through these surfaces (PECs) shown in the subplots of  Figure~\ref{Fig:1a1-2V-1_PECs}.
 
\section{Concluding remarks}
We  have computed the potential-energy surfaces for ions up to H$_2$O$^{3+}$ by means of the CASSCF \textit{ab initio} technique employing two active spaces and correlation-consistent quadruple-zeta basis sets.  Also, we identified whether the CASSCF, SA-CASSCF, SA-TS-CASSCF or the TS-CASSCF technique is the one providing the best accuracy depending on the water ion. To benchmark our results, we computed the  
 vertical ionization potentials with respect to the ground state of neutral water. We compared our results for the ionization potentials with state-of-the-art calculations and measurements available from the literature  
 for the singly and doubly charged ions as well as for triply charged ions with three electrons missing from valence electrons or with at least one electron missing from the 2a$_1$ orbital. We find  overall very good agreement with the available literature. This agreement suggests that the  PESs we compute for the triply charged ions with at least one electron missing from the 1a$_1$ orbital are also accurate. No other studies are available for these latter ion states.  We find that an interesting feature of the PECs as a function of the bond angle is that several states of the same symmetry undergo avoided crossings, while most PECs as a function of the bond distance are repulsive. Finally, the \textit{ab initio} approaches presented in this work provide a rationale for the computation of PESs for triatomic ions with an increasing number of electrons missing.
\section{Acknowledgements}
The authors acknowledge the use of the UCL Myriad High Performance Computing Facility (Myriad@UCL), and associated support services, in the completion of this work. A.E. is grateful for support by the Leverhulme Trust through the grant RPG-2025-179.
\bibliography{mainAE-2}
\end{document}